\documentclass[namedreferences]{solarphysics}

\usepackage[hyperref,optionalrh,showbiblabels]{spr-sola-addons} 
\usepackage{hyperref}
\hypersetup{
    colorlinks=true,
    linkcolor=red,
    filecolor=magenta,      
    urlcolor=cyan,
}

\usepackage{graphicx}        
\usepackage{color}           
\usepackage{breakurl}        

\chardef\us=`\_

\begin{document}

\begin{article}
\begin{opening}

\title{Exploring the circular polarisation of low--frequency solar radio bursts with LOFAR\\ {\it Solar Physics}}

\author[addressref={aff1},corref,email={diana.morosan@helsinki.fi}]{\inits{D. E.}\fnm{Diana E.}~\lnm{Morosan}}
\author[addressref=aff1,corref]{\inits{J. E.}\fnm{Juska E.}~\lnm{R{\"a}s{\"a}nen}}
\author[addressref=aff1,corref]{\inits{A.}\fnm{Anshu}~\lnm{Kumari}}
\author[addressref=aff1,corref]{\inits{E. K. J.}\fnm{Emilia K. J.}~\lnm{Kilpua}}

\author[addressref=aff2]{Mario M. Bisi}
\author[addressref=aff3]{Bartosz Dabrowski}
\author[addressref=aff3]{Andrzej Krankowski}
\author[addressref=aff4]{Jasmina Magdaleni\'{c}}
\author[addressref=aff5]{Gottfried Mann}
\author[addressref=aff6]{Hanna Rothkaehl}
\author[addressref=aff5]{Christian Vocks}
\author[addressref=aff7]{Pietro Zucca}

\address[id=aff1]{Department of Physics, University of Helsinki, P.O. Box 64, FI-00014 Helsinki, Finland}
\address[id=aff2]{RAL Space, United Kingdom Research and Innovation -- Science and Technology Facilities Council -- Rutherford
Appleton Laboratory, Harwell Campus, Oxfordshire OX11 0QX, UK }
\address[id=aff3]{Space Radio-Diagnostics Research Centre, University of Warmia and Mazury, Olsztyn, Poland}
\address[id=aff4]{Solar--Terrestrial Centre of Excellence---SIDC, Royal Observatory of Belgium, 1180 Brussels, Belgium}
\address[id=aff5]{Leibniz--Institut f\"{u}r Astrophysik Potsdam (AIP), An der Sternwarte 16, 14482, Potsdam, Germany}
\address[id=aff6]{Space Research Center, Polish Academy of Sciences, 00-716, Warsaw, Poland}
\address[id=aff7]{ASTRON, The Netherlands Institute for Radio Astronomy,
		Oude Hoogeveensedijk 4, 7991 PD Dwingeloo, The Netherlands}

\runningauthor{Morosan et al.}
\runningtitle{Exploring the circular polarisation of low--frequency solar radio bursts}

\begin{abstract}
The Sun is an active star that often produces numerous bursts of electromagnetic radiation at radio wavelengths. Low frequency radio bursts have recently been brought back to light with the advancement of novel radio interferometers. However, their polarisation properties have not yet been explored in detail, especially with the Low Frequency Array (LOFAR), due to difficulties in calibrating the data and accounting for instrumental leakage. Here, using a unique method to correct the polarisation observations, we explore the circular polarisation of different sub-types of solar type III radio bursts and a type I noise storm observed with LOFAR, which occurred during March--April 2019. We analysed six individual radio bursts from two different dates. We present the first Stokes V low frequency images of the Sun with LOFAR in tied-array mode observations. We find that the degree of circular polarisation for each of the selected bursts increases with frequency for fundamental emission, while this trend is either not clear or absent for harmonic emission. The type III bursts studied, that are part of a long--lasting type III storm, can have different senses of circular polarisation, occur at different locations and have different propagation directions. This indicates that the type III bursts forming a classical type III storm do not necessarily have a common origin but instead they indicate the existence of multiple, possibly unrelated, acceleration processes originating from solar minimum active regions.

\end{abstract}
\keywords{Polarization, Radio; Radio Bursts, Meter-Wavelenghts and Longer (m, dkm, hm, km), Type I, Type III; Radio Emission}
\end{opening}

\section{Introduction}
     \label{S-Introduction} 

{Bursts of electromagnetic radiation at radio wavelengths are a common phenomenon on the Sun. They can occur during large solar explosions and eruptions, namely flares and coronal mass ejections (CMEs), and also in their absence.  These bursts are classified into five main types: type I--V \citep{wild63}. Type I and type III radio bursts usually dominate the observations of solar activity at meter wavelengths in the absence of large eruptive events. }

{Type III bursts are rapidly varying bursts of radiation at meter wavelengths that show a fast drift in the dynamic spectra from high to low frequencies and last for a few seconds \citep[e.g.][]{wild50, Reid2014, mo14, mcc18, Mugundhan2018b}. Type III bursts are often associated with flaring activity, however they frequently occur in the absence of solar flares or other eruptive events \citep{du85}. They have been observed over a wide frequency range from GHz \citep[e.g.][]{ma12} to kHz \citep[e.g][]{kr18}, though most frequently, they occur at frequencies $<150$~MHz \citep{sa13}. Type III bursts represent the radio signature of electron beams travelling through the corona and into interplanetary space along open and quasi-open magnetic field lines \citep{Ginzburg1958,lin74, Reid2014}. Some sub-types of type III bursts, such as J- or U-bursts, indicate the presence of electron beams travelling along closed magnetic loops \citep{ma58, st71, mo17, re17}. It is commonly believed that following acceleration, faster electrons outpace the slower ones to produce a bump-on-tail instability in their velocity distribution. This generates Langmuir (plasma) waves which are then converted into radio waves at the plasma frequency, $f_p$, and its harmonic, $2f_p$ \citep[][]{ba98}, where the plasma frequency is defined as $f_p \sim C\sqrt{(n_e)}$, with $n_e$ being the electron number density and $C$ is a constant. }

{Type I bursts are non-thermal radio emissions generally associated with active regions \citep{mccready1947, Wild1951, Swarup1960, mel75}. They consist of short duration narrow--band bursts which can last for a few seconds to a few minutes \citep{takakura1963, Ellis1969, Mugundhan2018a}. These narrow--band bursts form a continuum emission known as a noise storm, which lasts for a few hours to a few days \citep{Elg1977, Thejappa1991, Kathiravan2007}. Type I noise storms/bursts are often found to be highly circularly polarised (see for example, \cite{Payne1949, Ramesh2013} and references therein). These bursts occur over a wide frequency range of $\sim30-200$ MHz \citep{Elg1977,Spicer1982, Yu2019}, with a few exceptions. }

{While the spectroscopic and spatial properties of type III and type I bursts have been studied in great detail \citep[e.g.][]{wi67, re95, sa13, mo14, ra20, Elg1977, Yu2019}, the circular polarisation of low frequency solar radio bursts is still poorly documented due to the previous unavailability of spectropolarimetric observations, except for a few instances, see for example \citet[][]{Payne1949, Elg1977, Ramesh2013, Mugundhan2018b}. Early observations of type III radio bursts have shown that the fundamental emission is highly circularly polarised, while the harmonic has a low degree of circular polarisation or it is unpolarised \citep{du80}. Type I noise storms are also highly polarised (up to $\sim100\%$) and they are believed to be generated at the fundamental plasma frequency\citep{Benz1985, Ramesh2013}. Recent studies with the Murchison Widefield Array \citep[MWA;][]{mwa13} produced spectropolarimetric images of the radio Sun at low frequencies in the range 80--240~MHz \citep{mcc19, rahman20}. Type III radio bursts have been studied in detail with the MWA, and \citet{rahman20} found that the degree of circular polarisation increases with frequency for fundamental emission type III bursts. However, no such studies have been carried out yet with LOFAR which is capable of observations at even lower frequencies (down to 20 MHz) compared to the MWA (down to 80~MHz). }

{Polarisation observations of solar radio bursts are still difficult to carry out due to the complexity in the calibration of polarised signals that are in part due to the lack of circularly polarised calibration sources and in part due to instrumental effects. With the advancement of novel radio instrumentation such as the Low Frequency Array \citep[LOFAR;][]{lofar13} and the MWA, spectropolarimetric imaging observations are now possible in the low frequency radio domain with a high temporal and frequency resolution. However, a careful investigation of instrumental effects and signal leakage must be taken into account. }

{In this paper, we present for the first time a method of obtaining the polarization information of solar radio bursts with LOFAR which takes into account instrumental leakage. We present the first spectropolarimetric and imaging observations with LOFAR of a few subtypes of type III bursts and a type I noise storm. In Section \ref{sec:section2} we give an overview of the LOFAR instrument, the observational mode used, and the data processing methods. In Section \ref{sec:section3} we present our results, which are further discussed in Section \ref{sec:section4}.}

\section{Observations and data analysis} 
\label{sec:section2}

\subsection{LOFAR full Stokes observations} 

{LOFAR is a next-generation radio telescope consisting of thousands of dipole antennas forming individual stations across Europe \citep{lofar13}. Here, we use tied-array beam observations of the Sun \citep{st11, lofar13} carried out during an observing campaign in March--Arpil 2019. These observations were recorded in the frequency range 20--80~MHz, using the Low Band Antennas (LBAs) from LOFAR's core, which is located in the Netherlands. We used a total of 126 beams that are pointing in a concentric pattern around the Sun to sample different locations. Due to the nature of the observations, and in order to accommodate the interferometric mode simultaneously with the tied-array beam observations, only a limited selection of frequency subbands were available for the beam-formed observations. The tied-array beam observations contain a total of 60 subbands distributed over 20--80 MHz. Each subband has a bandwidth of 195.3~kHz and is further divided into 16 frequency channels. The frequency resolution of each subband is 12.2~kHz. The frequency subbands are not uniformly distributed in the 20--80~MHz frequency range, therefore the frequency coverage is not continuous. However, due to the large number of subbands used, we can obtain an almost complete picture of the time--frequency evolution of individual radio bursts.}

{The tied-array beam observations are recorded as dynamic spectra in Stokes I, Q, U, and V for each of the 126 beams. The dynamic spectra are then calibrated using a separate beam pointing at an external calibrator throughout the observation. In this study, we used Taurus A as a calibrator. Stokes I, Q, U and V are given by the following relations \citep{Hales2017,Robishaw2021}:}


{
\begin{equation}  \label{eq1}
\label{eq:equation1}
    I = |A_x|^2 + |A_y|^2 
\end{equation}
\begin{equation}  \label{eq1}
\label{eq:equation1}
    Q = |A_x|^2 - |A_y|^2 
\end{equation}
\begin{equation}  \label{eq3}
\label{eq:equation1}
    U = 2 A_x A_y \mathrm{cos}\delta 
\end{equation}
\begin{equation}  \label{eq4}
\label{eq:equation2}
    V = 2 A_x A_y \mathrm{sin}\delta 
\end{equation}
where $A_x$ and $A_y$ are the dipole voltages from each of the two orthogonal LBA antennas and $\delta$ is the phase difference. For circular polarisation, $\delta = \pm \pi/2$.}

\begin{figure}[!ht]    
   \centerline{\includegraphics[width=1.6\linewidth,clip=]{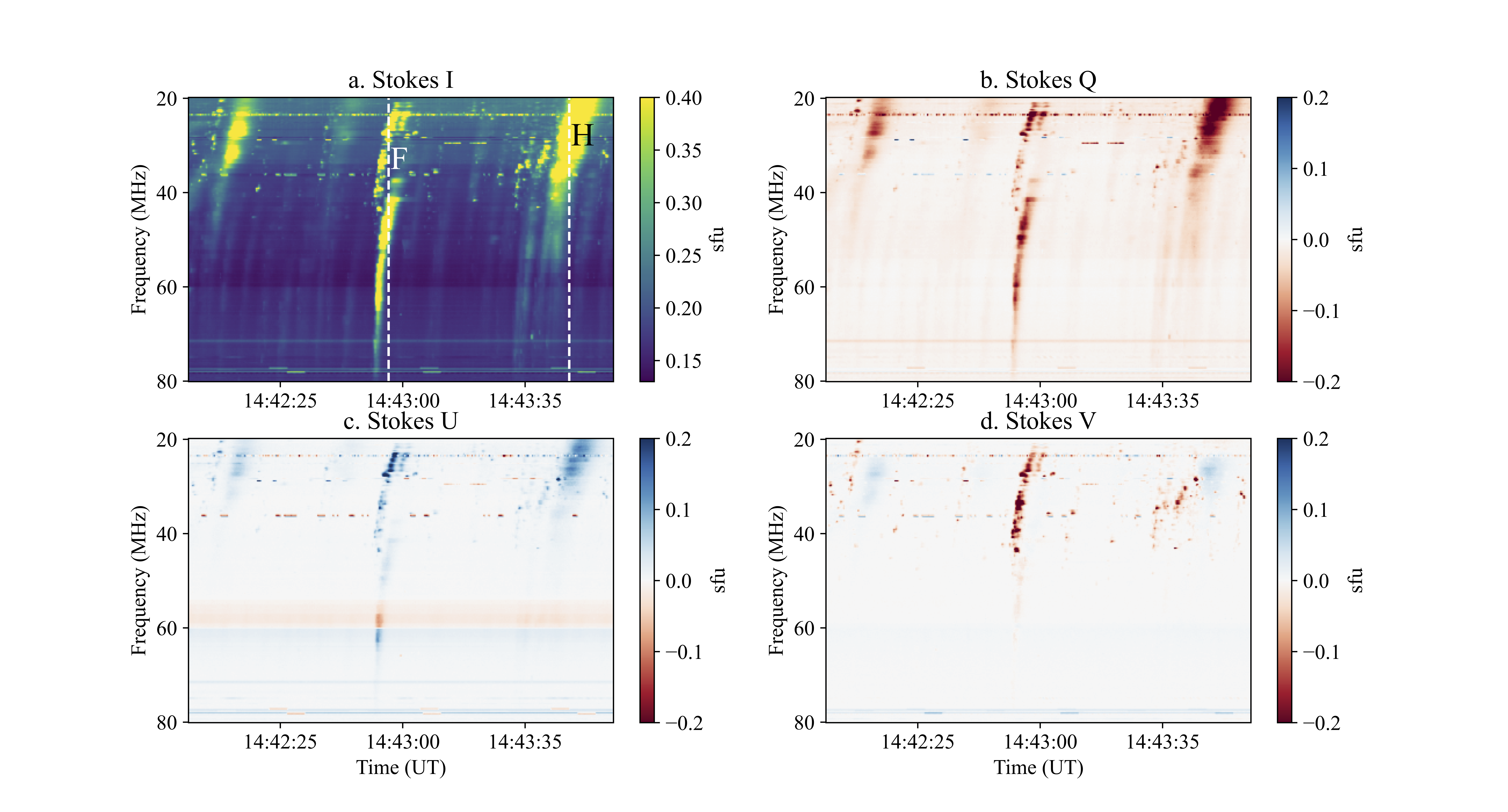}
              }
   \caption{Dynamic spectrum showing a number of type III radio bursts including both fundamental and harmonic emissions in Stokes I (a), Q (b), U (c) and V (d). }
   \label{fig1}
\end{figure}

{The dynamic spectrum beams are then used to make tied-array images of the Sun for each Stokes component using the methods of \citet{mo14, mo15}. This is done by interpolating between the intensity of emission from each individual beam, at a given time and frequency. The images made used the calibrated dynamic spectra and show flux density values in solar flux units (sfu, where 1 sfu $=10^{-22}$~W~m$^2$~Hz$^{-1}$) in both Stokes I and V. Each image is averaged over a time period of 0.5~s and over one full frequency subband. An example of a calibrated dynamic spectrum as well as tied-array images examples in Stokes I, Q, U and V (before correcting for instrumental leakage) are shown in Figs.~1--3 for all Stokes components.}

{At low radio frequencies ($<100$ MHz), the distance between the source and the observer is so large that the linear polarisation when averaged over an observing frequency band is expected to cancel out due to the Faraday rotation of the plane \citep{Grognard1973,Sasi2013}. Thus, the signals in Stokes Q and U (linearly polarised) is cancelled out over a certain size bandwidth. Therefore, the resulting low frequency solar radio emission is, in theory, only observed in Stokes I and V \citep[see for example][]{Sastry2009, kumari2017a, kumari2019}. However, some signal is often found in Stokes Q and U which most likely is leakage from Stokes I and Stokes V, due to the cross-correlation of the dipole antenna voltages \citep{Ramesh2008, kumari2017b, Mugundhan2018a, Kumari2021a}. This leakage is difficult to estimate due to the unavailability of polarised calibrators in the sky. }

\begin{figure}    
   \centerline{\includegraphics[width=\linewidth,clip=]{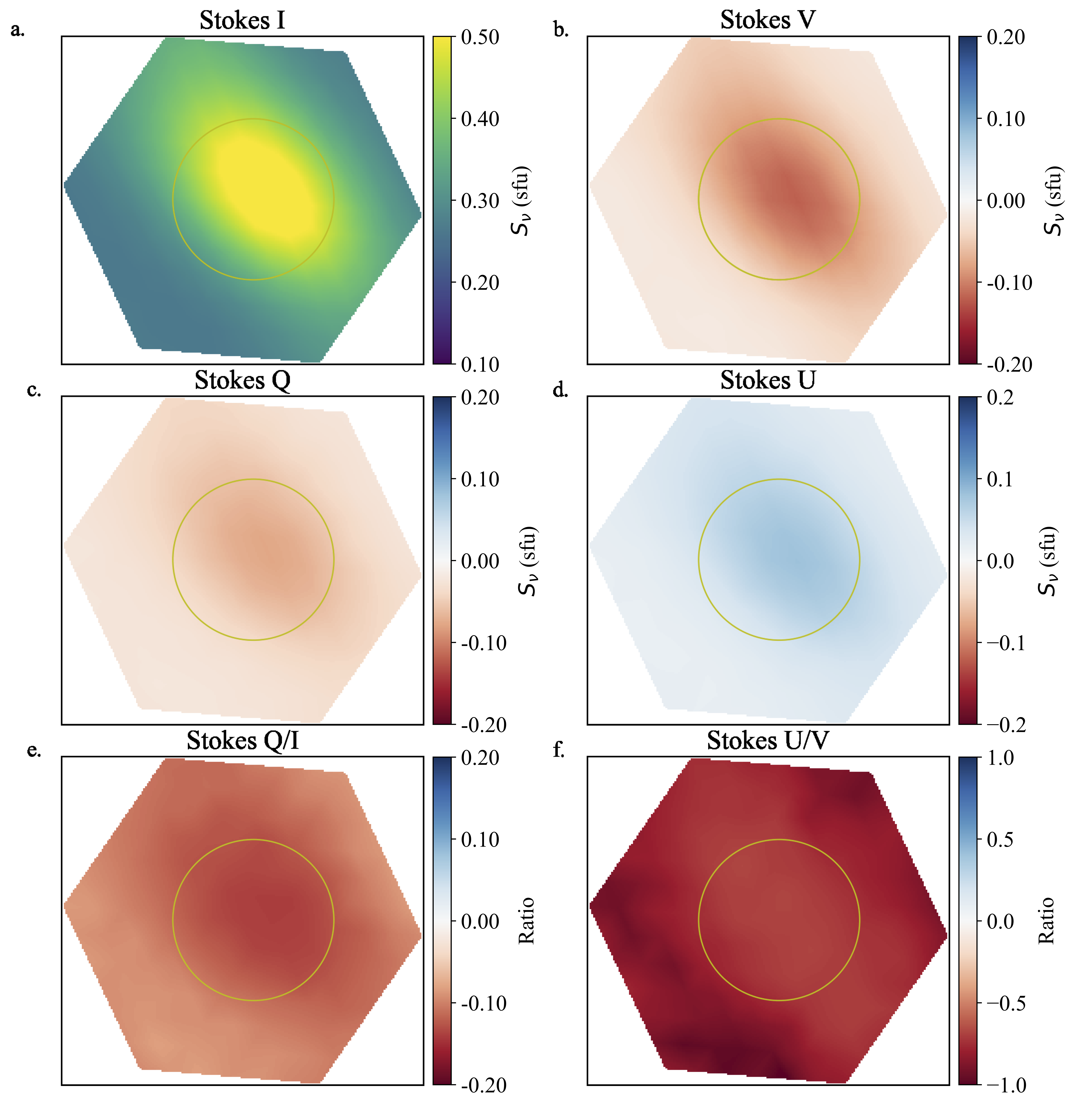}
              }
   \caption{LOFAR tied-array images of the type III radio burst denoted by the first dashed line in Fig.1 at 14:42:56~UT and a frequency of 27.4~MHz. Panels a--d show the Stokes I, Q, U and V images while the last two panels e--f show the ratios Q/I and U/V used to investigate leakage. In this case the signal is Stokes U is smaller than that in Stokes V. }
   \label{fig2}
\end{figure}

{The tied-array beams mode offers an unique possibility of quantifying the extent of polarisation leakage in Stokes Q and U, in order to obtain more accurate estimates of the signals in Stokes I and V. The signal can be analyzed in all 126 individual beams, both on and off source, to determine the extent of the leakage. However, this is done under the assumption that no residual linearly polarised signal is detected. A dynamic spectrum of a few type III radio bursts is shown in Fig.~\ref{fig1} in Stokes I (a), Q (b), U (c) and V (d). While there is clear signal in Stokes V which shows the circular polarisation of some of these type III bursts, significant signal is also observed in Stokes Q and U. For the fundamental type III radio burst (labelled as F in Fig.~\ref{fig1}), we observe significant signal in Stokes V and some weaker signal in Stokes U which would indicate leakage from Stokes V to Stokes U. However, for the radio bursts that are weak in Stokes V, which are mostly harmonic emissions (labelled as H in Fig.~\ref{fig1}), the signal in Stokes U is greater than that in Stokes V. All radio bursts show signal in Stokes Q which indicates leakage from Stokes I into Q. }

\begin{figure}    
   \centerline{\includegraphics[width=\linewidth,clip=]{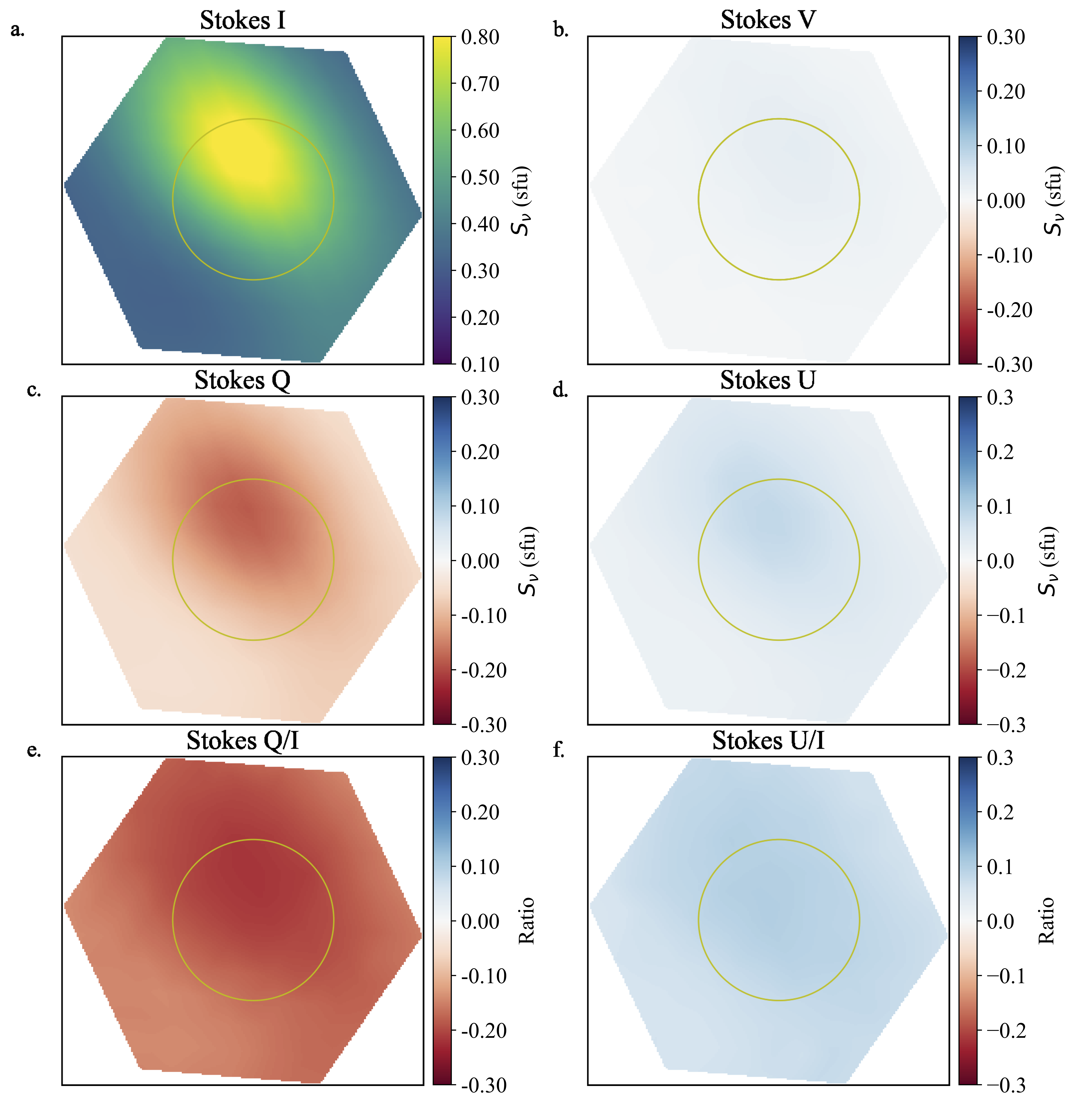}
              }
   \caption{LOFAR tied-array images of the type III radio burst denoted by the second dashed line in Fig.1 at 14:43:47~UT and a frequency of 27~MHz. Panels a--d show the Stokes I, Q, U and V images while the last two panels e--f show the ratios Q/I and U/I used to investigate leakage. In this case there is no circularly polarised signal and thus the signal is Stokes U is greater than that in Stokes V. }
   \label{fig3}
\end{figure}

{In order to quantify the amount of leakage in Stokes Q and U and determine the origin of the additional signal in Stokes U when there is no signal in Stokes V, we investigated tied-array images in all Stokes components. Thus, we can estimate the polarisation leakage on and off source (in our case in radio burst beams and quiet beams, respectively). We investigated two cases where the signal in Stokes U is smaller than that in Stokes V (Fig.~\ref{fig2}) and the signal in Stokes U is greater than that in Stokes V (Fig.~\ref{fig3}). Then, by taking ratios $U/I$, $U/V$, $Q/I$, we can determine the percentage of leakage of I and V into either Stokes Q and U. When there is signal in Stokes V (Fig.~\ref{fig2}), there is on-source leakage from Stokes V to U, however, in the beams pointing off-source, the ratio U/V $>$ 1 (Fig.~\ref{fig2}f) which indicates that not all signal leakage is from Stokes V. When there is no significant signal in Stokes V (Fig.~\ref{fig3}b) and thus U $>$ V (Fig.~\ref{fig3}d), there is signal in U both on and off-source, which is consistent with leakage from Stokes I, since the ratio U/I (Fig.~\ref{fig3}f) is roughly uniform across all beams and shows less variation compared to the ratio Q/I. There is also continuous leakage from Stokes I into Stokes Q both on and off source. We did not find any clear evidence of leakage of Stokes V into Q. This is hard to identify as Stokes I is always present and dominates the emission observed in Stokes Q. }

{The Stokes I spectra ($I_{corr}$) are thus corrected in the following way:
\begin{equation}  \label{eq3}
\label{eq:equation3}
     I_{corr} = I + F_Q I + F_U I,
\end{equation}
where $F_Q$ and $F_U$ represent the fraction of Stokes I leakage into Stokes Q and U, respectively. The leakage in Stokes U is approximately uniform and a value of $\sim$5\% was selected which represents the off-source leakage. The signal in Stokes Q can vary with subband and on- and off-source positions and it is in the range 5--30\% of the signal in Stokes I. We selected the median value of $F_Q$ for each subband to correct Stokes I.} 

{The Stokes V spectra ($V_{corr}$) are corrected using the following relation, to include the signal leakage in U back into the Stokes V component, as follows:
\begin{equation}  \label{eq3}
\label{eq:equation3}
     V_{corr} = \frac{V}{|V|}\sqrt{V^2+(|U|-F_U I)^2}.
\end{equation}
Here, the ratio $V/|V|$ is used to preserve the sign of V, since correcting Stokes V requires a sum of squares method. The sign of V is also strongly dependent on the sign of $\delta$, unlike Stokes U. The square root term of the equation adds back the Stokes V signal spilled into Stokes U after Stokes U is corrected for leakage from Stokes I. The leakage of I into U is small ($F_U=\sim$5\%) and is subtracted from the absolute value of U so that it is not added into Stokes V. The leakage of V into U is in the range of 5--30\% depending on on- and off--source location and subband. }

\subsection{Radio bursts observations} 

{During the LOFAR observing campaign in March--April 2019, we observed a long--lasting type III storm, including numerous sub--types of type III bursts, and also a long--lasting type I noise storm. The observed bursts are not associated with a particular solar flare, however a few C-class flares occurred at the end of March 2019 and only some B-class flares in the beginning of April 2019, when the bursts presented in this study were observed.}

{In this study, we investigate the circular polarisation and spatial location of six individual radio bursts outlined in Table~\ref{tab1}, in order to demonstrate LOFAR's spectropolarimetric capabilities. For each burst we extracted the time series averaged over a full subband in all Stokes components. We then used Stokes I, Q, U and V to obtain the degree of circular polarisation of each individual burst. Stokes V is used to obtain information on the sense of circular polarisation of each burst. Fig.~\ref{fig4} shows the dynamic spectra for these individual bursts along with the time series of the subband--averaged Stokes components. The non-uniform subband coverage often results in a not so smooth transition between frequency channels which can make the harmonic components of type III bursts appear patchy in a similar way to fundamental emission (see for example the harmonic emission intersected by the first white dashed vertical line in Fig.~\ref{fig1}a). }

\begin{table}    
\label{T-simple}
\begin{tabular}{ccclc}     
  \hline                   

Burst & Date and start time (UT) & Type & Maximum dcp & Sense \\
  \hline
1 & 20190321 14:22:59 & FH Type III Pair  & 65\% (F)/15\% (H)~ & -V \\
2 & 20190321 14:33:27 & Stria chain  & 45\% (F) & -V \\
3 & 20190321 14:42:50 & FH Type III Pair & 70\% (F)/20\% (H) & -V \\
4 & 20190321 14:44:42 & J-burst Pair  & 15\% (H) & +V \\
5 & 20190408 12:20:00 & Type I & 100\% (F) & -V \\
6 & 20190408 12:58:43 &FH Type III Pair & 50\% (F)/20\% (H) & -V \\
  \hline
  
\end{tabular}
\caption{ Radio bursts observed with LOFAR and their properties including date and time, type (FH=Fundamental-Harmonic), maximum degree of circular polarisation (dcp) estimated in Section 3.3, and sense of circular polarisation.}\label{tab1}
\end{table}

\section{Results} 
\label{sec:section3}

\subsection{Radio bursts and their circular polarisation}

\begin{figure}    
   \centerline{\includegraphics[width=1.3\linewidth,clip=]{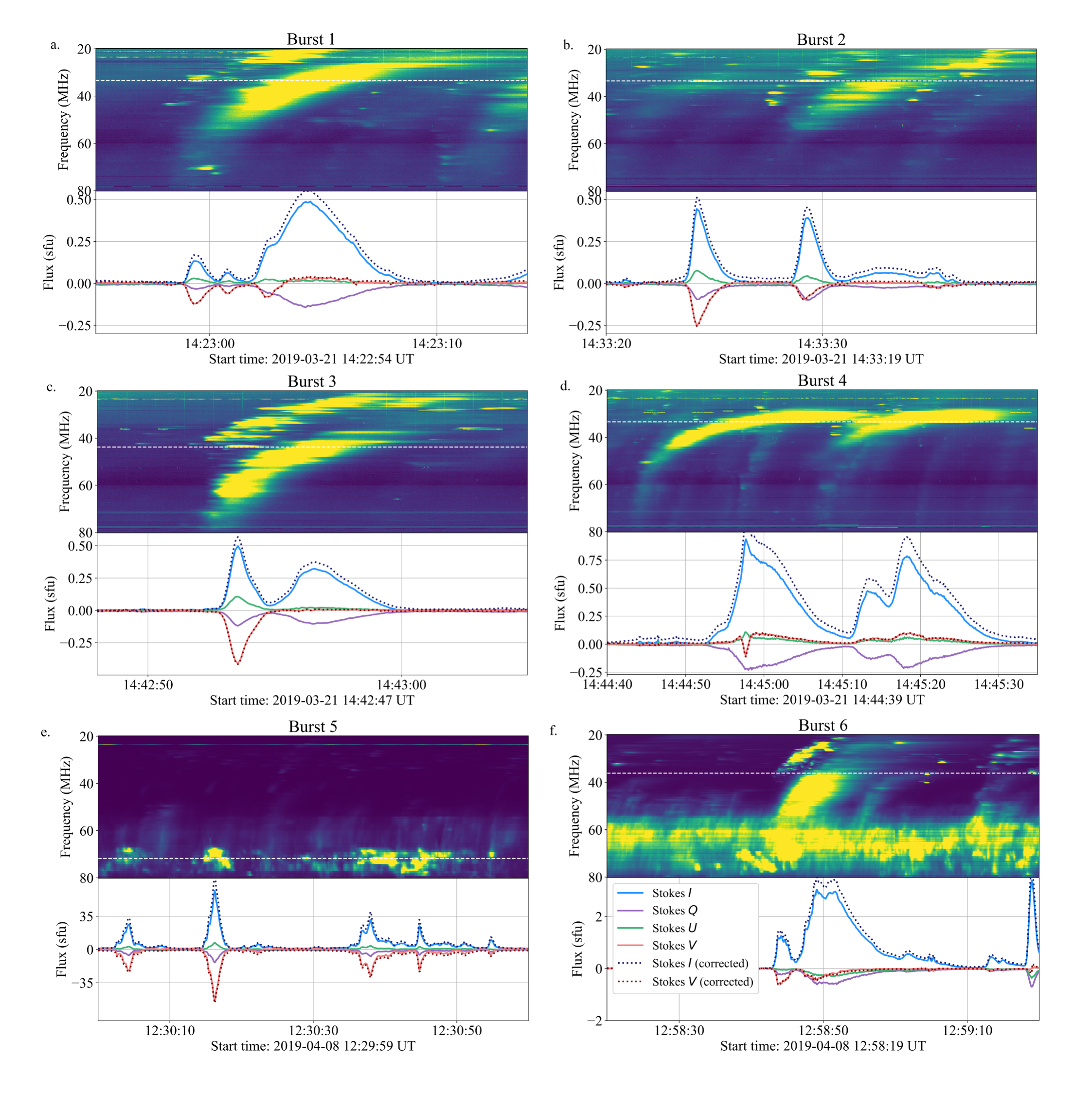}
              }
    \caption{Dynamic spectra in Stokes I and time series for Stokes I (\textit{`blue'}), Q (\textit{`purple'}), U (\textit{`green'}), V (\textit{`red'}), I corrected (\textit{`dark blue dotted'}) and V corrected (\textit{`dark red dotted'}) for the bursts outlined in Table~\ref{tab1}. For each burst, the time series is averaged over the subband outlined by the white dashed lines in the dynamic spectra with a center a frequency of 31~MHz (a, b, d, and f), 43~MHz (c), and, 70~MHz (f).}
    \label{fig4}
\end{figure}

The radio bursts outlined in Table~\ref{tab1} are studied in detail to obtain their degree of circular polarisation (dcp) and sense of circular polarisation (determined using the sign of V). Table~\ref{tab1} also contains a summary of their properties such as date and start time, type, maximum dcp and, sense of circular polarisation. The radio bursts include sub-types of type III radio bursts (such as stria bursts and J-bursts) and a type I noise storm (Burst 5). Bursts 1--4 were observed on 21 March 2019 and Bursts 5--6 on 8 April 2019.  The spectral and polarisation properties these bursts are discussed here:

{\textbf{Burst 1} consists of a type III fundamental--harmonic (FH) pair observed on 21 March 2019 which started at 14:22:59~UT (Fig.~\ref{fig4}a). The FH components extend over the entire frequency range of the observation. An example of a time--series averaged over one subband extending from 31.34--31.53~MHz and a centre frequency of 31.44~MHz is shown in the bottom panel of Fig.~\ref{fig4}a. The time series reveals that the fundamental emission (which corresponds to the stria burst occurring before 14:23~UT) appears highly polarised i.e. the Stokes V intensity (\textit{`red solid'} line) is comparable to that of Stokes I (\textit{`blue solid'} line). On the other hand, the harmonic (which corresponds to the smoother burst after 14:23~UT) appears almost unpolarised (Stokes V$\sim$0) which agrees with previous findings \citep[e.g.][]{du80}. This is consistent for all the individual subbands in the 20--80~MHz range. The sign of Stokes V is negative. There is however a small signal in Stokes U (\textit{`green solid'} line in Fig.~\ref{fig4}a) which is used to correct the Stokes V signal and significant signal in Stokes Q used to correct the Stokes I signal.}

{\textbf{Burst 2} represents a chain of stria bursts (the intermittent structures usually forming a fundamental type III burst) observed on 21 March 2019 starting at 14:33:27~UT (Fig.~\ref{fig4}b). The stria chain extends from 20--50~MHz. The stria bursts usually represent emission forming a fundamental type III burst and are moderately circularly polarised (Fig.~\ref{fig4}b) which remains consistent throughout each subband containing the bursts. The sign of Stokes V is negative. }

{\textbf{Burst 3} consists of another type III FH pair observed on 21 March 2019 starting at 14:42:50~UT (Fig.~\ref{fig2}c). In this case the fundamental component is also highly circularly polarised while the harmonic is almost unpolarised. The sign of Stokes V is negative similar to the previous FH type III burst discussed. }

{\textbf{Burst 4} consists of a pair of J-bursts observed on 21 March 2019, which started at 14:44:42~UT (Fig.~\ref{fig2}d). The J-bursts represent harmonic emission that is weakly circularly polarised. The sign of Stokes V is positive, which is opposite to the other bursts studied here. Since J-burst emitting electrons propagate along closed magnetic field lines, it is likely that they propagate along different magnetic field lines to the other bursts observed on 21 March 2019 that are expected to propagate along the open field instead. Thus, they can have a different sense of circular polarisation to the other bursts.  }

{\textbf{Burst 5} consists of type I bursts forming a long-lasting type I noise storm that started on 8 April 2019 at $\sim$12:30~UT (Fig.~\ref{fig2}e). The type I bursts are strongly circularly polarised and occur simultaneously with fainter type III radio bursts.  The sign of Stokes V is negative. The type I emission is only observed in the 60--80~MHz frequency range. }

{\textbf{Burst 6} consists of a type III FH pair observed on 4 April 2019 starting at 12:58:43~UT (Fig.~\ref{fig2}f), during the type I noise storm. This type III was selected as it has a similar flux density value to the type I noise storm and it also occurs on a background of fainter type III activity. The fundamental component is also highly circularly polarised while the harmonic is weakly polarised. The sign of Stokes V is negative, same as with the type I noise storm. This burst has a higher flux density than the other FH bursts reported here.  }

\subsection{Degree of circular polarisation of observed bursts}

\begin{figure}    
   \centerline{\includegraphics[width=1.3\linewidth,clip=]{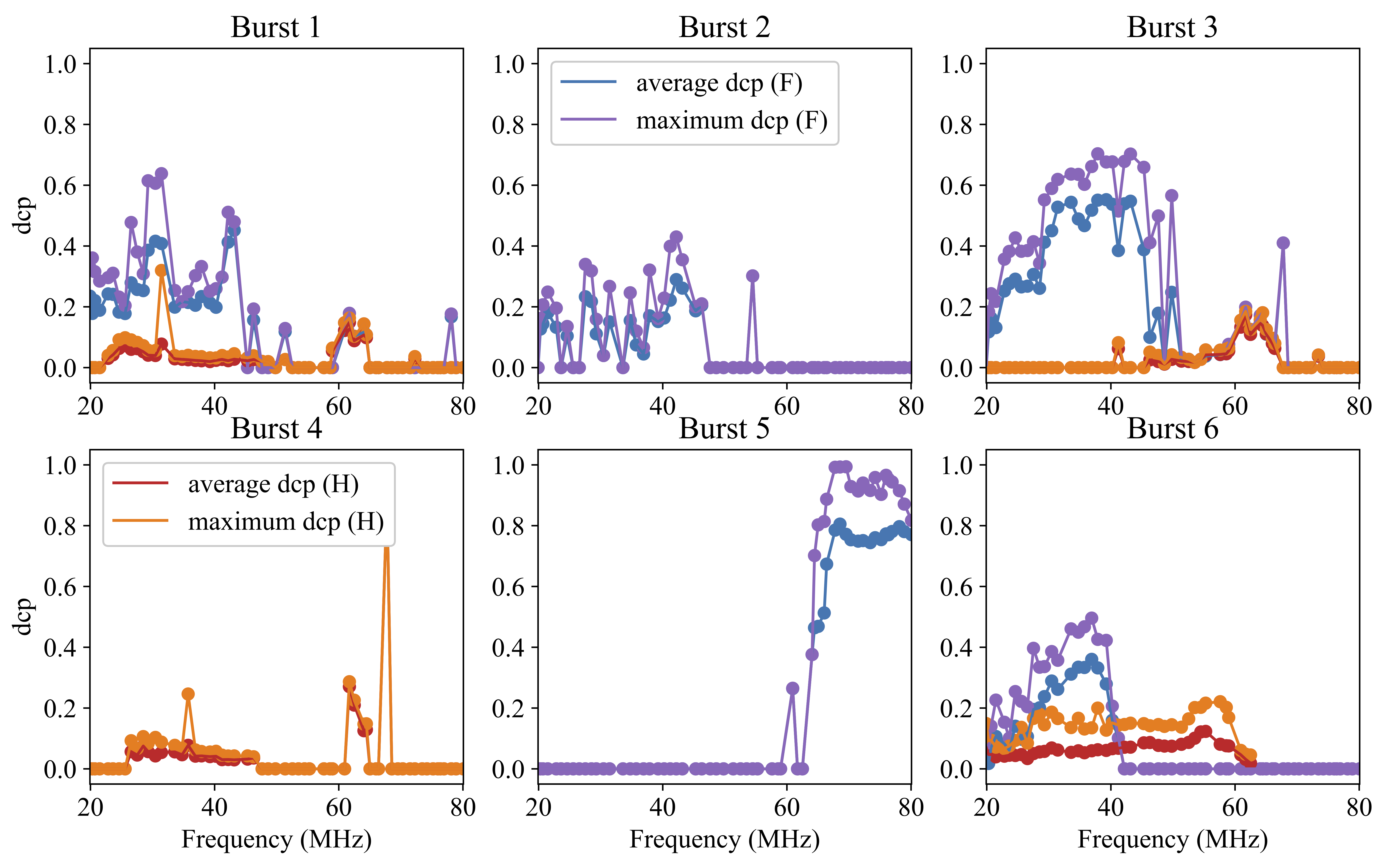}
              }
   \caption{Degree of circular polarisation (dcp) of the bursts outlined in Table~\ref{tab1} as a function of frequency. The \textit{`red', `orange', `blue', and `purple'} lines represent the average and maximum dcp of harmonic (H) and fundamental (F) components of the bursts discussed here, respectively. }
   \label{fig5}
\end{figure}

{For each of the bursts outlined in the previous section, we computed the degree of circular polarisation (dcp) as a function of frequency (Fig.~\ref{fig3}). The degree of circular polarisation (dcp) is obtained as, 
\begin{equation}
    \mathrm{dcp} = \frac{V_{corr}}{I_{corr}}
\end{equation}
The dynamic spectra and radio images can now be used to extract simultaneously the degree of circular polarisation, location, and trajectory of the radio emission that are presented in the following sections. The dcp was computed by averaging the data over a time period of 0.5~s and over a full frequency subband. The maximum dcp is summarised in Table~\ref{tab1}. The panels in Fig.~\ref{fig5} show the average and maximum dcp for each subband in the case of each burst. The average and maximum dcp were computed for each subband over a time range comprising of the entire duration of each burst. }

{We found that the dcp for type III fundamental emission is large (40--75\%) while the observed harmomnic emission is weakly polarised (up to 25\%). In the case of the bursts studied here, the dcp is higher than the average dcp of type III bursts recorded by \citet{du80} in the range of 24--220 MHz, but comparable to values reported in more recent studies by \citet{rahman20}. The type I storm is also highly circularly polarised with dcp values reaching 70\%, in agreement with previous studies \citep[e.g.][]{du73}. We also found a clear trend of the dcp increasing with frequency in the case of fundamental emissions consistent with the findings of \citet[][]{rahman20} with the MWA. This is also seen in the case of harmonic emission for Bursts 3 and 6, however, an opposite trend is seen for Bursts 1 and 4. }

{In the case of Bursts 1, 2 and 5, there appear to be fluctuations in both the average and maximum dcp values. However, due to the uneven distribution of subbands in this observational setup, fine-structured bursts, especially striae, can be abruptly cut off. These fluctuations arise due to the discontinuous frequency coverage. In the case of the fundamental emission for Bursts 1 and 2, fluctuations may also arise due to averaging of the signal over the individual striae or the gap in between the striae. }

\subsection{Locations of observed bursts and their sense of circular polarisation}

\begin{figure}    
   \centerline{\includegraphics[width=1.3\linewidth,clip=]{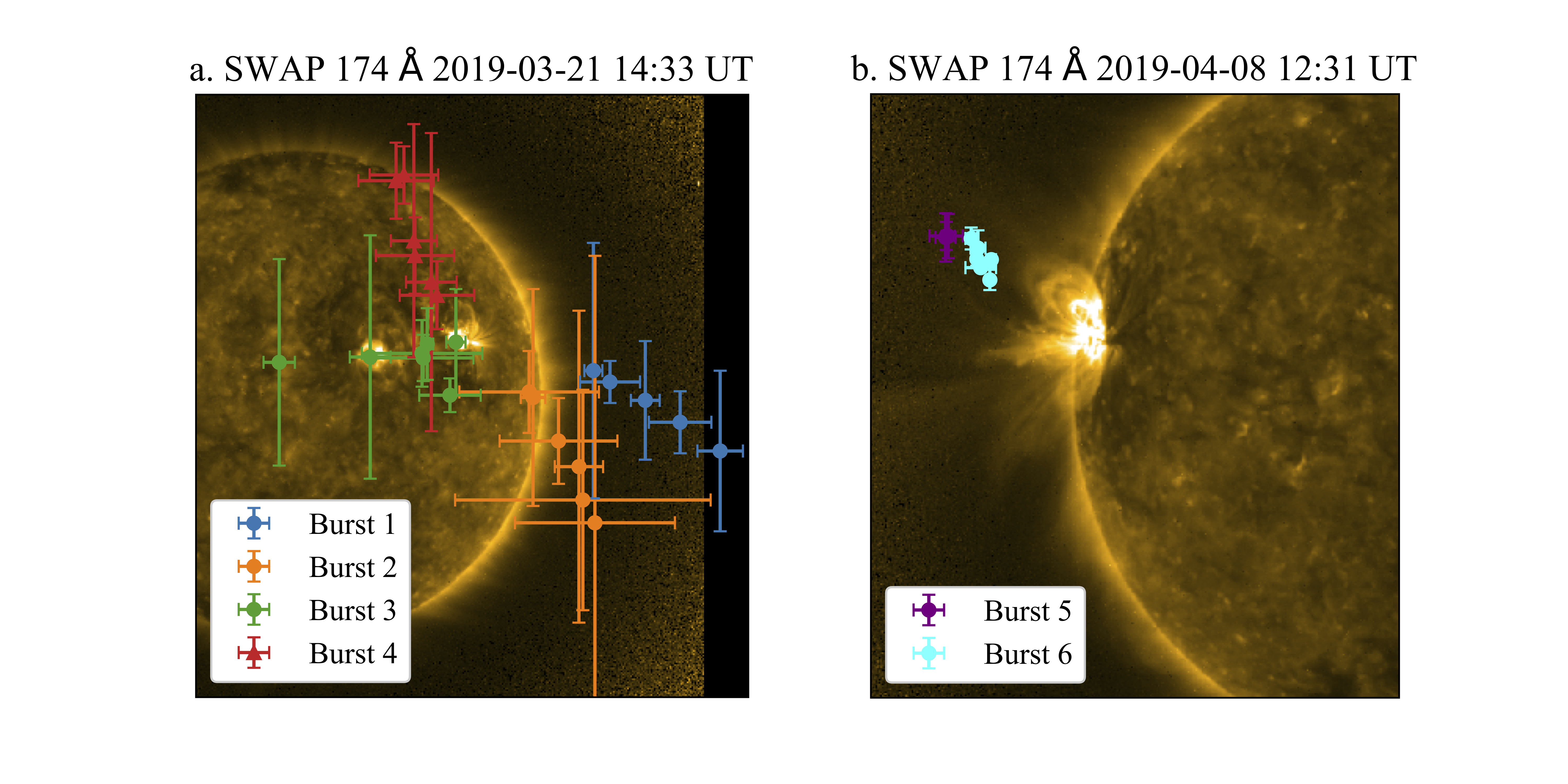}
              }
   \caption{Radio burst centroids obtained from tied-array images of the Sun. (a) Centroids of the type III bursts (Bursts 1--4) observed on 21 March 2019 overlaid on a PROBA-2/SWAP extreme-ultraviolet image of the Sun. (b) Centroids of the type I and type III bursts (Bursts 5 and 6, respectively) observed on 4 April 2020 overlaid on a PROBA-2/SWAP extreme-ultraviolet image of the Sun. The centroids are obtained from images at several frequencies spread over the spectral extent of each burst. Burst 1 centroids are at frequencies of 29-47~MHz, Burst 2 at 20-41~MHz, Burst 3 at 24-43~MHz, Burst 4 at 30-43~MHz, Burst 5 at 24-55~MHz, and Burst 6 at frequencies of 61-76~MHz. }
   \label{fig6}
\end{figure}

{The radio centroids of the observed bursts were extracted from the tied-array images by fitting a two-dimensional elliptical Gaussian to the radio sources (for more information, see the methods described in \citealt{mo19}). The radio centroids are shown in the panels of Fig.~\ref{fig6} overlaid on top of extreme ultraviolet images from the Sun-Watcher with Active Pixel System and Image Processing \citep[SWAP;][]{se13} instrument onboard the Project for On Board Autonomy 2 \citep[PROBA2;][]{Santandrea2013} spacecraft. The centroids are obtained from images at different frequencies spread over the spectral extent of each burst, in order to show the propagation direction of the bursts. Burst 1 centroids are at frequencies of 29-47~MHz, Burst 2 at 20-41~MHz, Burst 3 at 24-43~MHz, Burst 4 at 30-43~MHz, Burst 5 at 24-55~MHz, and Burst 6 at frequencies of 61-76~MHz.  }

{The location of the centroids show that radio bursts that occurred on 21 March 2019 have different propagation directions which is in agreement with the wide variety of dcp values and different senses of circular polarisation observed. It is also interesting to note that Burst 1 and Burst 3, that occur on 21 Mach 2019 and are both FH type III bursts, originate at different locations and propagate in opposite directions as seen in Fig.~\ref{fig6}. The FH pair in Burst 1 propagates radially outwards, while the FH pair in Burst 3 propagates inwards across the solar disk. The J-bursts (Burst 4) are propagating towards the North pole most likely along close magnetic loops, with one footpoint in one of the active regions present in the western hemisphere. A scenario that could explain these propagation directions is open or quasi-open magnetic field lines branching outwards as a fan and electron beams can then escape along these different field lines. All bursts appear to originate from the solar hemisphere containing active regions and most likely originate due to processes in or near the two active regions present in the western hemisphere.  }

{The centroids of the bursts on 8 April 2019 (both type I and type III) have smaller uncertainties due to the much larger flux densities of these bursts compared to the background emission \citep[for more details in estimating the centroid uncertainties please see][]{mo19}. The overlap of various frequencies in the same spatial location indicates that despite being limb events, plane-of-sky projection effects are significant in the case of the these bursts. The type III, in particular, is propagating out of the plane-of-sky towards or away from the observer. In this case, it is more likely that these bursts are propagating away from the observer since the active region is only just rotating on the visible solar disc and the magnetic field lines are mostly connected to regions still behind the disc.  }

\section{Discussion and conclusion} 
\label{sec:section4}
    
{Type III and type I radio bursts are emitted by plasma radiation in the solar corona. This emission originates as a result of Langmuir (plasma) waves that are generated by accelerated electron beams propagating through the coronal plasma \citep{ Stewart1972,Suzuki1985}. For Langmuir waves travelling parallel or anti-parallel to the magnetic field, the polarisation is in the sense of the ordinary (o-) mode \citep{du76, du80}. The o-mode is the only mode that can theoretically propagate through the coronal plasma to produce fundamental plasma emission, since the x-mode is blocked \citep{me09}. Fundamental radiation is then expected to be up to 100\% polarised \citep{me72}. Type III and type I bursts have also been often reported as o-mode radiation \citep[e.g.][]{du80, du73}. }

{In our study, the FH type III bursts observed on 21 March 2019 (Burst 1 and Burst 3) show fundamental emission that is highly circularly polarised (65\% and 75\%, respectively). Fundamental emission is expected to be up to 100\% polarised, however previous studies reported that type III bursts show clear depolarisation effects since the observed dcp $<<100\%$ \citep{du80, mercier90, rahman20}. On the other hand, observations of type I storms have found that they are up to 100\% circularly polarised \citep[e.g.][]{kai85,mu18}. The type I burst observed here, Burst 5, is indeed up to 100\% circularly polarized. The main theory attributed to the depolarisation of type III bursts is due to scattering effects of radio waves as they propagate through the coronal plasma \citep{ro82, me89,rahman20}. In the case of Bursts 1 and 3 observed here, we report very high dcp values for fundamental emission, especially during the burst onset at higher frequencies, however, the highest dcp value is 75\% similar to the previous studies discussed. The higher dcp at burst onset implies that propagation effects in the solar corona may become more significant with decreasing frequency since the dcp value also decreases with frequency. Since Bursts 1 and 3 have higher dcp values than the other type III bursts and all bursts have different propagation directions, it seems that the significance of scattering of radio waves in the corona may be dependent on the location, propagation direction, and coronal conditions immediately in the vicinity of the radio source regions. }

{In the case of fundamental emission for the type III there is a clear trend of the dcp increasing with frequency. This trend is only clear in the case of Bursts 3 and 6 for harmonic emission and an opposite trend is seen for Bursts 1 and 4. In the case of harmonic emission, an increase of dcp with frequency occurs due to the fact that the dcp of radio bursts is proportional to the strength of the coronal magnetic field which decreases with height and thus with plasma frequency \citep{melrose80, mercier90}. The opposite trend in Bursts 1 and 4 cannot be explained by this relation with magnetic field and most likely there may still be some unaccounted leakage in Stoke U or V. The fundamental on the other hand is expected to be 100\% circularly polarised, therefore the reason for the existence of such a trend for fundamental emission is unclear. This trend in the fundamental has also been reported with the MWA \citep{rahman20}. Depolarisation within the source region has been suggested as a possible explanation \citep{wentzel84}. \citet{rahman20} show that beaming effects in the source region as a function of the magnetic field strength, which in turn is dependant on radial height, leads to the dcp to also decrease with height in the corona. In the case of the highly polarised Bursts 1 and 3, we find that the type III is highly circularly polarised during burst onset after which the dcp slowly decreases. In contrast, the Type I burst appears to have an almost constant dcp over each subband (Fig.~\ref{fig5}).}

{The most noteworthy finding is that all four type III bursts that occurred on 21 March 2019 within a time period of 30~minutes have a wide range of dcp values and one of these also shows an opposite sense of circular polarisation. This is explained by the fact that all of these bursts have different propagation directions. Strongly different source positions and propagation directions of subsequent type III radio bursts associated with the same eruptive event were also recently reported by \citet{jebaraj2020}. Different senses of circular polarisation within the same group or storm of type III bursts have previously been reported only in the case of type IIIs followed by type V bursts \citep[][]{dulk1980}. Type V bursts represent continuum emission following a type III burst or group of type III bursts that can show an opposite sense of circular polarisation to the preceding type III bursts \citep[][]{dulk1980}. However, recent observations of type V bursts with modern instrumentation have yet to be reported, which may have implications on the classification of these bursts as an entirely different type of bursts than type IIIs. Here, we have shown that it is possible to have different senses of polarisation within the same storm of type III bursts. Despite solar minimum conditions, multiple acceleration processes producing energetic electron beams occur in or in the vicinity of the two active regions that were present on the Sun on the day. The active regions appear to be the source regions for the observed bursts. The type III and type I occurring on 8 April 2019 appear to have a similar source region, that is the active region on the eastern solar limb. }

{This study demonstrates for the first time LOFAR's spectropolarimetric imaging capabilities to investigate solar radio bursts. We used full stokes parameters to estimate the correct degree of circular polarisation for type I and type III bursts which occurred during March--April 2019. Our analysis shows that LOFAR is suitable to be used as a dynamic spectropolarimetric imaging instrument to study radio emission from the Sun. }


\begin{acks}
D.E.M. acknowledges the Academy of Finland Project `RadioCME' (grant number 333859) and the Finnish Centre of Excellence in Research of Sustainable Space (Academy of Finland grant number 312390). J.E.R. acknowledges the University of Helsinki Faculty of Science Support Fund. A.K. and E.K.J.K. acknowledge the ERC under the European Union's Horizon 2020 Research and Innovation Programme Project SolMAG 724391. E.K.J.K. also acknowledges the Academy of Finland Project 310445. B.D. and A.K. thank the National Science Centre, Poland for granting “LOFAR observations of the solar corona during Parker Solar Probe perihelion passages” in the Beethoven Classic 3 funding initiative under project number 2018/31/G/ST9/01341. This paper is based (in part) on data obtained with the International LOFAR Telescope (ILT) under project codes LC11\_001 and LT10\_002. LOFAR (van Haarlem et al. 2013) is the Low Frequency Array designed and constructed by ASTRON. It has observing, data processing, and data storage facilities in several countries, that are owned by various parties (each with their own funding sources), and that are collectively operated by the ILT foundation under a joint scientific policy. The ILT resources have benefited from the following recent major funding sources: CNRS-INSU, Observatoire de Paris and Université d'Orléans, France; BMBF, MIWF-NRW, MPG, Germany; Science Foundation Ireland (SFI), Department of Business, Enterprise and Innovation (DBEI), Ireland; NWO, The Netherlands; The Science and Technology Facilities Council, UK, Ministry of Science and Higher Education, Poland. The authors would like to thank Michiel Brentjens and Maaijke Mevius for the useful discussion on LOFAR polarisation observations. 
\end{acks}

\section*{Declarations}
The LOFAR datasets used in this analysis were obtained under the project codes LC11\_001 and LT10\_002 and they are available in the LOFAR Long Term Archive (LTA; \url{https://lta.lofar.eu}). The SWAP data is available from the Virtual Solar Observatory project (\url{http://vso.nso.edu}).



\bibliographystyle{spr-mp-sola}
\bibliography{main_revisions_arxiv}

\begin{thebibliography}{70}
\ifx\bisbn     \undefined \def\bisbn  #1{ISBN #1}\fi
\ifx\binits    \undefined \def\binits#1{#1}\fi
\ifx\bauthor   \undefined \def\bauthor#1{#1}\fi
\ifx\batitle   \undefined \def\batitle#1{#1}\fi
\ifx\bjtitle   \undefined \def\bjtitle#1{\textit{#1}}\fi
\ifx\bvolume   \undefined \def\bvolume#1{\textbf{#1}}\fi
\ifx\byear     \undefined \def\byear#1{#1}\fi
\ifx\bissue    \undefined \def\bissue#1{#1}\fi
\ifx\bfpage    \undefined \def\bfpage#1{#1}\fi
\ifx\blpage    \undefined \def\blpage #1{#1}\fi
\ifx\burl      \undefined \def\burl#1{\textsf{#1}}\fi
\ifx\href      \undefined \def\href#1#2{\textsf{#2}}\fi
\ifx\betal     \undefined \def\betal{\textit{et al.}}\fi
\ifx\bctitle   \undefined \def\bctitle#1{#1}\fi
\ifx\beditor   \undefined \def\beditor#1{#1}\fi
\ifx\bbtitle   \undefined \def\bbtitle#1{\textit{#1}}\fi
\ifx\bedition  \undefined \def\bedition#1{#1}\fi
\ifx\bseriesno \undefined \def\bseriesno#1{\textbf{#1}}\fi
\ifx\blocation \undefined \def\blocation#1{#1}\fi
\ifx\bsertitle \undefined \def\bsertitle#1{\textit{#1}}\fi
\ifx\bsnm      \undefined \def\bsnm#1{#1}\fi
\ifx\bsuffix   \undefined \def\bsuffix#1{#1}\fi
\ifx\bparticle \undefined \def\bparticle#1{#1}\fi
\ifx\barticle  \undefined \def\barticle#1{}\fi
\ifx\binstitute  \undefined \def\binstitute#1{#1}\fi
\ifx\bpublisher  \undefined \def\bpublisher#1{#1}\fi
\ifx\doiurl    \undefined
  \def\doiurl#1{\href{http://dx.doi.org/#1}{\textsf{DOI}}}\fi
\ifx\arxivurl  \undefined
  \def\arxivurl#1{\href{http://arxiv.org/abs/#1}{\textsf{arXiv}}}\fi
\ifx\adsurl    \undefined
  \def\adsurl#1{\href{http://adsabs.harvard.edu/abs/#1}{\textsf{ADS}}}\fi
\ifx\botherref \undefined \def\botherref#1{}\fi
\ifx\url       \undefined \def\url#1{\textsf{#1}}\fi
\ifx\bchapter  \undefined \def\bchapter#1{}\fi
\ifx\bbook     \undefined \def\bbook#1{}\fi
\ifx\bcomment  \undefined \def\bcomment#1{#1}\fi
\ifx\oauthor   \undefined \def\oauthor#1{#1}\fi
\ifx\citeauthoryear \undefined\def \citeauthoryear#1{#1}\fi
\ifx\endbibitem\undefined \def\endbibitem{}\fi
\ifx\bconflocation  \undefined \def\bconflocation#1{#1} \fi

\bibitem[\protect\citeauthoryear{{Bastian}, {Benz}, and {Gary}}{1998}]{ba98}
\begin{barticle}
\bauthor{\bsnm{{Bastian}}, \binits{T.S.}},
\bauthor{\bsnm{{Benz}}, \binits{A.O.}},
\bauthor{\bsnm{{Gary}}, \binits{D.E.}}:
\byear{1998},
\batitle{{Radio Emission from Solar Flares}}.
\bjtitle{\araa}
\bvolume{36},
\bfpage{131}.
\doiurl{10.1146/annurev.astro.36.1.131}.
\adsurl{1998ARA\%26A..36..131B}.
\end{barticle}
\endbibitem

\bibitem[\protect\citeauthoryear{{Benz} and {Zolliker}}{1985}]{Benz1985}
\begin{barticle}
\bauthor{\bsnm{{Benz}}, \binits{A.O.}},
\bauthor{\bsnm{{Zolliker}}, \binits{P.}}:
\byear{1985},
\batitle{{Polarization of solar noise storm continuum and plasma wave density
  in the corona}}.
\bjtitle{\aap}
\bvolume{144}(\bissue{1}),
\bfpage{227}.
\adsurl{https://ui.adsabs.harvard.edu/abs/1985A&A...144..227B}.
\end{barticle}
\endbibitem

\bibitem[\protect\citeauthoryear{{Dulk}}{1985}]{du85}
\begin{barticle}
\bauthor{\bsnm{{Dulk}}, \binits{G.A.}}:
\byear{1985},
\batitle{{Radio emission from the sun and stars}}.
\bjtitle{\araa}
\bvolume{23},
\bfpage{169}.
\doiurl{10.1146/annurev.aa.23.090185.001125}.
\adsurl{1985ARA\%26A..23..169D}.
\end{barticle}
\endbibitem

\bibitem[\protect\citeauthoryear{{Dulk} and {Nelson}}{1973}]{du73}
\begin{barticle}
\bauthor{\bsnm{{Dulk}}, \binits{G.A.}},
\bauthor{\bsnm{{Nelson}}, \binits{G.J.}}:
\byear{1973},
\batitle{{The Position of a Type I Storm Source in the Magnetic Field of an
  Active Region}}.
\bjtitle{Proceedings of the Astronomical Society of Australia}
\bvolume{2}(\bissue{4}),
\bfpage{211}.
\doiurl{10.1017/S132335800001362X}.
\adsurl{https://ui.adsabs.harvard.edu/abs/1973PASAu...2..211D}.
\end{barticle}
\endbibitem

\bibitem[\protect\citeauthoryear{{Dulk} and {Suzuki}}{1980}]{du80}
\begin{barticle}
\bauthor{\bsnm{{Dulk}}, \binits{G.A.}},
\bauthor{\bsnm{{Suzuki}}, \binits{S.}}:
\byear{1980},
\batitle{{The position and polarization of Type III solar bursts}}.
\bjtitle{\aap}
\bvolume{88}(\bissue{1-2}),
\bfpage{203}.
\adsurl{https://ui.adsabs.harvard.edu/abs/1980A&A....88..203D}.
\end{barticle}
\endbibitem

\bibitem[\protect\citeauthoryear{{Dulk}, {Gary}, and {Suzuki}}{1980}]{dulk1980}
\begin{barticle}
\bauthor{\bsnm{{Dulk}}, \binits{G.A.}},
\bauthor{\bsnm{{Gary}}, \binits{D.E.}},
\bauthor{\bsnm{{Suzuki}}, \binits{S.}}:
\byear{1980},
\batitle{{The position and polarization of Type V solar bursts}}.
\bjtitle{\aap}
\bvolume{88}(\bissue{1-2}),
\bfpage{218}.
\adsurl{https://ui.adsabs.harvard.edu/abs/1980A&A....88..218D}.
\end{barticle}
\endbibitem

\bibitem[\protect\citeauthoryear{{Dulk} \textit{et~al.}}{1976}]{du76}
\begin{barticle}
\bauthor{\bsnm{{Dulk}}, \binits{G.A.}},
\bauthor{\bsnm{{Jacques}}, \binits{S.}},
\bauthor{\bsnm{{Smerd}}, \binits{S.F.}},
\bauthor{\bsnm{{MacQueen}}, \binits{R.M.}},
\bauthor{\bsnm{{Gosling}}, \binits{J.T.}},
\bauthor{\bsnm{{Steward}}, \binits{R.T.}},
\bauthor{\bsnm{{Sheridan}}, \binits{K.V.}},
\bauthor{\bsnm{{Robinson}}, \binits{R.D.}},
\bauthor{\bsnm{{Magun}}, \binits{A.}}:
\byear{1976},
\batitle{{White Light and Radio Studies of the Coronal Transient of 14-15
  September 1973. I: Material Motions and Magnetic Field}}.
\bjtitle{\solphys}
\bvolume{49},
\bfpage{369}.
\adsurl{https://ui.adsabs.harvard.edu/abs/1976SoPh...49..369D}.
\end{barticle}
\endbibitem

\bibitem[\protect\citeauthoryear{{Elgar{\o}y}}{1977}]{Elg1977}
\begin{bbook}
\bauthor{\bsnm{{Elgar{\o}y}}, \binits{E.{\O}.}}:
\byear{1977},
\bbtitle{{Solar noise storms.}}
\adsurl{https://ui.adsabs.harvard.edu/abs/1977sns..book.....E}.
\end{bbook}
\endbibitem

\bibitem[\protect\citeauthoryear{{Ellis}}{1969}]{Ellis1969}
\begin{barticle}
\bauthor{\bsnm{{Ellis}}, \binits{G.R.A.}}:
\byear{1969},
\batitle{{Fine structure in the spectra of solar radio bursts.}}
\bjtitle{Australian Journal of Physics}
\bvolume{22},
\bfpage{177}.
\doiurl{10.1071/PH690177}.
\adsurl{https://ui.adsabs.harvard.edu/abs/1969AuJPh..22..177E}.
\end{barticle}
\endbibitem

\bibitem[\protect\citeauthoryear{{Ginzburg} and
  {Zhelezniakov}}{1958}]{Ginzburg1958}
\begin{barticle}
\bauthor{\bsnm{{Ginzburg}}, \binits{V.L.}},
\bauthor{\bsnm{{Zhelezniakov}}, \binits{V.V.}}:
\byear{1958},
\batitle{{On the Possible Mechanisms of Sporadic Solar Radio Emission
  (Radiation in an Isotropic Plasma)}}.
\bjtitle{\sovast}
\bvolume{2},
\bfpage{653}.
\adsurl{https://ui.adsabs.harvard.edu/abs/1958SvA.....2..653G}.
\end{barticle}
\endbibitem

\bibitem[\protect\citeauthoryear{{Grognard} and {McLean}}{1973}]{Grognard1973}
\begin{barticle}
\bauthor{\bsnm{{Grognard}}, \binits{R.J.M.}},
\bauthor{\bsnm{{McLean}}, \binits{D.J.}}:
\byear{1973},
\batitle{{Non-Existence of Linear Polarization in Type III Solar Bursts at 80
  MHz}}.
\bjtitle{\solphys}
\bvolume{29}(\bissue{1}),
\bfpage{149}.
\doiurl{10.1007/BF00153446}.
\adsurl{https://ui.adsabs.harvard.edu/abs/1973SoPh...29..149G}.
\end{barticle}
\endbibitem

\bibitem[\protect\citeauthoryear{{Hales}}{2017}]{Hales2017}
\begin{barticle}
\bauthor{\bsnm{{Hales}}, \binits{C.A.}}:
\byear{2017},
\batitle{{Calibration Errors in Interferometric Radio Polarimetry}}.
\bjtitle{\aj}
\bvolume{154}(\bissue{2}),
\bfpage{54}.
\doiurl{10.3847/1538-3881/aa7aef}.
\adsurl{https://ui.adsabs.harvard.edu/abs/2017AJ....154...54H}.
\end{barticle}
\endbibitem

\bibitem[\protect\citeauthoryear{{Jebaraj} \textit{et~al.}}{2020}]{jebaraj2020}
\begin{barticle}
\bauthor{\bsnm{{Jebaraj}}, \binits{I.C.}},
\bauthor{\bsnm{{Magdaleni{\'c}}}, \binits{J.}},
\bauthor{\bsnm{{Podladchikova}}, \binits{T.}},
\bauthor{\bsnm{{Scolini}}, \binits{C.}},
\bauthor{\bsnm{{Pomoell}}, \binits{J.}},
\bauthor{\bsnm{{Veronig}}, \binits{A.M.}},
\bauthor{\bsnm{{Dissauer}}, \binits{K.}},
\bauthor{\bsnm{{Krupar}}, \binits{V.}},
\bauthor{\bsnm{{Kilpua}}, \binits{E.K.J.}},
\bauthor{\bsnm{{Poedts}}, \binits{S.}}:
\byear{2020},
\batitle{{Using radio triangulation to understand the origin of two subsequent
  type II radio bursts}}.
\bjtitle{\aap}
\bvolume{639},
\bfpage{A56}.
\doiurl{10.1051/0004-6361/201937273}.
\adsurl{https://ui.adsabs.harvard.edu/abs/2020A&A...639A..56J}.
\end{barticle}
\endbibitem

\bibitem[\protect\citeauthoryear{{Kai}, {Melrose}, and {Suzuki}}{1985}]{kai85}
\begin{bbook}
\bauthor{\bsnm{{Kai}}, \binits{K.}},
\bauthor{\bsnm{{Melrose}}, \binits{D.B.}},
\bauthor{\bsnm{{Suzuki}}, \binits{S.}}:
\byear{1985},
In: \beditor{\bsnm{{McLean}}, \binits{D.J.}},
\beditor{\bsnm{{Labrum}}, \binits{N.R.}} (eds.)
\bbtitle{{Storms.}},
\bfpage{415}.
\adsurl{https://ui.adsabs.harvard.edu/abs/1985srph.book..415K}.
\end{bbook}
\endbibitem

\bibitem[\protect\citeauthoryear{{Kathiravan}, {Ramesh}, and
  {Nataraj}}{2007}]{Kathiravan2007}
\begin{barticle}
\bauthor{\bsnm{{Kathiravan}}, \binits{C.}},
\bauthor{\bsnm{{Ramesh}}, \binits{R.}},
\bauthor{\bsnm{{Nataraj}}, \binits{H.S.}}:
\byear{2007},
\batitle{{The Post-Coronal Mass Ejection Solar Atmosphere and Radio Noise Storm
  Activity}}.
\bjtitle{\apjl}
\bvolume{656}(\bissue{1}),
\bfpage{L37}.
\doiurl{10.1086/512013}.
\adsurl{https://ui.adsabs.harvard.edu/abs/2007ApJ...656L..37K}.
\end{barticle}
\endbibitem

\bibitem[\protect\citeauthoryear{Krupar \textit{et~al.}}{2018}]{kr18}
\begin{barticle}
\bauthor{\bsnm{Krupar}, \binits{V.}},
\bauthor{\bsnm{Maksimovic}, \binits{M.}},
\bauthor{\bsnm{Kontar}, \binits{E.P.}},
\bauthor{\bsnm{Zaslavsky}, \binits{A.}},
\bauthor{\bsnm{Santolik}, \binits{O.}},
\bauthor{\bsnm{Soucek}, \binits{J.}},
\bauthor{\bsnm{Kruparova}, \binits{O.}},
\bauthor{\bsnm{Eastwood}, \binits{J.P.}},
\bauthor{\bsnm{Szabo}, \binits{A.}}:
\byear{2018},
\batitle{Interplanetary type {III} bursts and electron density fluctuations in
  the solar wind}.
\bjtitle{The Astrophysical Journal}
\bvolume{857}(\bissue{2}),
\bfpage{82}.
\doiurl{10.3847/1538-4357/aab60f}.
\burl{https://doi.org/10.3847\%2F1538-4357\%2Faab60f}.
\end{barticle}
\endbibitem

\bibitem[\protect\citeauthoryear{{Kumari} \textit{et~al.}}{2017a}]{kumari2017a}
\begin{barticle}
\bauthor{\bsnm{{Kumari}}, \binits{A.}},
\bauthor{\bsnm{{Ramesh}}, \binits{R.}},
\bauthor{\bsnm{{Kathiravan}}, \binits{C.}},
\bauthor{\bsnm{{Gopalswamy}}, \binits{N.}}:
\byear{2017}a,
\batitle{{New Evidence for a Coronal Mass Ejection-driven High Frequency Type
  II Burst near the Sun}}.
\bjtitle{\apj}
\bvolume{843},
\bfpage{10}.
\doiurl{10.3847/1538-4357/aa72e7}.
\adsurl{2017ApJ...843...10K}.
\end{barticle}
\endbibitem

\bibitem[\protect\citeauthoryear{{Kumari} \textit{et~al.}}{2017b}]{kumari2017b}
\begin{barticle}
\bauthor{\bsnm{{Kumari}}, \binits{A.}},
\bauthor{\bsnm{{Ramesh}}, \binits{R.}},
\bauthor{\bsnm{{Kathiravan}}, \binits{C.}},
\bauthor{\bsnm{{Wang}}, \binits{T.J.}}:
\byear{2017}b,
\batitle{{Strength of the Solar Coronal Magnetic Field - A Comparison of
  Independent Estimates Using Contemporaneous Radio and White-Light
  Observations}}.
\bjtitle{Solar Physics}
\bvolume{292}(\bissue{11}),
\bfpage{161}.
\doiurl{10.1007/s11207-017-1180-6}.
\adsurl{https://ui.adsabs.harvard.edu/abs/2017SoPh..292..161K}.
\end{barticle}
\endbibitem

\bibitem[\protect\citeauthoryear{{Kumari} \textit{et~al.}}{2019}]{kumari2019}
\begin{barticle}
\bauthor{\bsnm{{Kumari}}, \binits{A.}},
\bauthor{\bsnm{{Ramesh}}, \binits{R.}},
\bauthor{\bsnm{{Kathiravan}}, \binits{C.}},
\bauthor{\bsnm{{Wang}}, \binits{T.J.}},
\bauthor{\bsnm{{Gopalswamy}}, \binits{N.}}:
\byear{2019},
\batitle{{Direct Estimates of the Solar Coronal Magnetic Field Using
  Contemporaneous Extreme-ultraviolet, Radio, and White-light Observations}}.
\bjtitle{The Astrophysical Journal}
\bvolume{881}(\bissue{1}),
\bfpage{24}.
\doiurl{10.3847/1538-4357/ab2adf}.
\adsurl{https://ui.adsabs.harvard.edu/abs/2019ApJ...881...24K}.
\end{barticle}
\endbibitem

\bibitem[\protect\citeauthoryear{{Kumari} \textit{et~al.}}{2021}]{Kumari2021a}
\begin{botherref}
\oauthor{\bsnm{{Kumari}}, \binits{A.}},
\oauthor{\bsnm{{Gireesh}}, \binits{G.V.S.}},
\oauthor{\bsnm{{Kathiravan}}, \binits{C.}},
\oauthor{\bsnm{{Mugundhan}}, \binits{V.}},
\oauthor{\bsnm{{Barve}}, \binits{I.V.}}:
2021,
{Solar Radio Spectro-polarimetry (50-500 MHz) : Design and Development of
  Cross-Polarized Log-Periodic Dipole antenna and configuration of receiver
  system}.
\textit{arXiv e-prints},
arXiv:2101.05088.
\adsurl{https://ui.adsabs.harvard.edu/abs/2021arXiv210105088K}.
\end{botherref}
\endbibitem

\bibitem[\protect\citeauthoryear{{Lin}}{1974}]{lin74}
\begin{barticle}
\bauthor{\bsnm{{Lin}}, \binits{R.P.}}:
\byear{1974},
\batitle{{Non-relativistic Solar Electrons}}.
\bjtitle{\ssr}
\bvolume{16},
\bfpage{189}.
\doiurl{10.1007/BF00240886}.
\adsurl{1974SSRv...16..189L}.
\end{barticle}
\endbibitem

\bibitem[\protect\citeauthoryear{{Ma} \textit{et~al.}}{2012}]{ma12}
\begin{barticle}
\bauthor{\bsnm{{Ma}}, \binits{Y.}},
\bauthor{\bsnm{{Xie}}, \binits{R.-x.}},
\bauthor{\bsnm{{Zheng}}, \binits{X.-m.}},
\bauthor{\bsnm{{Wang}}, \binits{M.}},
\bauthor{\bsnm{{Yi-hua}}, \binits{Y.}}:
\byear{2012},
\batitle{{A Statistical Analysis of the Type-III Bursts in Centimeter and
  Decimeter Wavebands}}.
\bjtitle{\caa}
\bvolume{36},
\bfpage{175}.
\doiurl{10.1016/j.chinastron.2012.04.008}.
\adsurl{2012ChA\%26A..36..175M}.
\end{barticle}
\endbibitem

\bibitem[\protect\citeauthoryear{{Maxwell} and {Swarup}}{1958}]{ma58}
\begin{barticle}
\bauthor{\bsnm{{Maxwell}}, \binits{A.}},
\bauthor{\bsnm{{Swarup}}, \binits{G.}}:
\byear{1958},
\batitle{{A New Spectral Characteristic in Solar Radio Emission}}.
\bjtitle{\nat}
\bvolume{181},
\bfpage{36}.
\doiurl{10.1038/181036a0}.
\adsurl{1958Natur.181...36M}.
\end{barticle}
\endbibitem

\bibitem[\protect\citeauthoryear{{McCauley}, {Cairns}, and
  {Morgan}}{2018}]{mcc18}
\begin{barticle}
\bauthor{\bsnm{{McCauley}}, \binits{P.I.}},
\bauthor{\bsnm{{Cairns}}, \binits{I.H.}},
\bauthor{\bsnm{{Morgan}}, \binits{J.}}:
\byear{2018},
\batitle{{Densities Probed by Coronal Type III Radio Burst Imaging}}.
\bjtitle{\solphys}
\bvolume{293}(\bissue{10}),
\bfpage{132}.
\doiurl{10.1007/s11207-018-1353-y}.
\adsurl{https://ui.adsabs.harvard.edu/abs/2018SoPh..293..132M}.
\end{barticle}
\endbibitem

\bibitem[\protect\citeauthoryear{{McCauley} \textit{et~al.}}{2019}]{mcc19}
\begin{barticle}
\bauthor{\bsnm{{McCauley}}, \binits{P.I.}},
\bauthor{\bsnm{{Cairns}}, \binits{I.H.}},
\bauthor{\bsnm{{White}}, \binits{S.M.}},
\bauthor{\bsnm{{Mondal}}, \binits{S.}},
\bauthor{\bsnm{{Lenc}}, \binits{E.}},
\bauthor{\bsnm{{Morgan}}, \binits{J.}},
\bauthor{\bsnm{{Oberoi}}, \binits{D.}}:
\byear{2019},
\batitle{{The Low-Frequency Solar Corona in Circular Polarization}}.
\bjtitle{\solphys}
\bvolume{294}(\bissue{8}),
\bfpage{106}.
\doiurl{10.1007/s11207-019-1502-y}.
\adsurl{https://ui.adsabs.harvard.edu/abs/2019SoPh..294..106M}.
\end{barticle}
\endbibitem

\bibitem[\protect\citeauthoryear{McCready, Pawsey, and
  Payne-Scott}{1947}]{mccready1947}
\begin{barticle}
\bauthor{\bsnm{McCready}, \binits{L.}},
\bauthor{\bsnm{Pawsey}, \binits{J.L.}},
\bauthor{\bsnm{Payne-Scott}, \binits{R.}}:
\byear{1947},
\batitle{Solar radiation at radio frequencies and its relation to sunspots}.
\bjtitle{Proceedings of the Royal Society of London. Series A. Mathematical and
  Physical Sciences}
\bvolume{190}(\bissue{1022}),
\bfpage{357}.
\end{barticle}
\endbibitem

\bibitem[\protect\citeauthoryear{{Melrose}}{1975}]{mel75}
\begin{barticle}
\bauthor{\bsnm{{Melrose}}, \binits{D.B.}}:
\byear{1975},
\batitle{{Plasma emission due to isotropic fast electrons, and types I, II, and
  V solar radio bursts}}.
\bjtitle{\solphys}
\bvolume{43},
\bfpage{211}.
\doiurl{10.1007/BF00155154}.
\adsurl{1975SoPh...43..211M}.
\end{barticle}
\endbibitem

\bibitem[\protect\citeauthoryear{{Melrose}}{1989}]{me89}
\begin{barticle}
\bauthor{\bsnm{{Melrose}}, \binits{D.B.}}:
\byear{1989},
\batitle{{Depolarization of Solar Bursts due to Scattering by Low-Frequency
  Waves}}.
\bjtitle{\solphys}
\bvolume{119}(\bissue{1}),
\bfpage{143}.
\doiurl{10.1007/BF00146218}.
\adsurl{https://ui.adsabs.harvard.edu/abs/1989SoPh..119..143M}.
\end{barticle}
\endbibitem

\bibitem[\protect\citeauthoryear{{Melrose}}{2009}]{me09}
\begin{bchapter}
\bauthor{\bsnm{{Melrose}}, \binits{D.B.}}:
\byear{2009},
\bctitle{{Coherent emission}}.
In: \beditor{\bsnm{{Gopalswamy}}, \binits{N.}},
\beditor{\bsnm{{Webb}}, \binits{D.F.}} (eds.)
\bbtitle{Universal Heliophysical Processes},
\bsertitle{IAU Symposium}
\bseriesno{257},
\bfpage{305}.
\doiurl{10.1017/S1743921309029470}.
\adsurl{2009IAUS..257..305M}.
\end{bchapter}
\endbibitem

\bibitem[\protect\citeauthoryear{{Melrose} and {Sy}}{1972}]{me72}
\begin{barticle}
\bauthor{\bsnm{{Melrose}}, \binits{D.B.}},
\bauthor{\bsnm{{Sy}}, \binits{W.N.}}:
\byear{1972},
\batitle{{Plasma emission processes in a magnetoactive plasma}}.
\bjtitle{Aust. J. Phys.}
\bvolume{25},
\bfpage{387}.
\doiurl{10.1071/PH720387}.
\adsurl{https://ui.adsabs.harvard.edu/abs/1972AuJPh..25..387M}.
\end{barticle}
\endbibitem

\bibitem[\protect\citeauthoryear{{Melrose}, {Dulk}, and
  {Gary}}{1980}]{melrose80}
\begin{barticle}
\bauthor{\bsnm{{Melrose}}, \binits{D.B.}},
\bauthor{\bsnm{{Dulk}}, \binits{G.A.}},
\bauthor{\bsnm{{Gary}}, \binits{D.E.}}:
\byear{1980},
\batitle{{Corrected formula for the polarization of second harmonic plasma
  emission}}.
\bjtitle{Proceedings of the Astronomical Society of Australia}
\bvolume{4}(\bissue{1}),
\bfpage{50}.
\adsurl{https://ui.adsabs.harvard.edu/abs/1980PASAu...4...50M}.
\end{barticle}
\endbibitem

\bibitem[\protect\citeauthoryear{{Mercier}}{1990}]{mercier90}
\begin{barticle}
\bauthor{\bsnm{{Mercier}}, \binits{C.}}:
\byear{1990},
\batitle{{Polarisation of Type-Iii Bursts Between 164-MHZ and 435-MHZ -
  Structure and Variation with Frequency}}.
\bjtitle{\solphys}
\bvolume{130}(\bissue{1-2}),
\bfpage{119}.
\doiurl{10.1007/BF00156783}.
\adsurl{https://ui.adsabs.harvard.edu/abs/1990SoPh..130..119M}.
\end{barticle}
\endbibitem

\bibitem[\protect\citeauthoryear{{Morosan} \textit{et~al.}}{2014}]{mo14}
\begin{barticle}
\bauthor{\bsnm{{Morosan}}, \binits{D.E.}},
\bauthor{\bsnm{{Gallagher}}, \binits{P.T.}},
\bauthor{\bsnm{{Zucca}}, \binits{P.}},
\bauthor{\bsnm{{Fallows}}, \binits{R.}},
\bauthor{\bsnm{{Carley}}, \binits{E.P.}},
\bauthor{\bsnm{{Mann}}, \binits{G.}},
\bauthor{\bparticle{et.} \bsnm{al.}}:
\byear{2014},
\batitle{{LOFAR tied-array imaging of Type III solar radio bursts}}.
\bjtitle{\aap}
\bvolume{568},
\bfpage{A67}.
\doiurl{10.1051/0004-6361/201423936}.
\adsurl{2014A\%26A...568A..67M}.
\end{barticle}
\endbibitem

\bibitem[\protect\citeauthoryear{{Morosan} \textit{et~al.}}{2015}]{mo15}
\begin{barticle}
\bauthor{\bsnm{{Morosan}}, \binits{D.E.}},
\bauthor{\bsnm{{Gallagher}}, \binits{P.T.}},
\bauthor{\bsnm{{Zucca}}, \binits{P.}},
\bauthor{\bsnm{{O'Flannagain}}, \binits{A.}},
\bauthor{\bsnm{{Fallows}}, \binits{R.}},
\bauthor{\bsnm{{Reid}}, \binits{H.}},
\bauthor{\bparticle{et.} \bsnm{al.}}:
\byear{2015},
\batitle{{LOFAR tied-array imaging and spectroscopy of solar S bursts}}.
\bjtitle{\aap}
\bvolume{580},
\bfpage{A65}.
\doiurl{10.1051/0004-6361/201526064}.
\adsurl{https://ui.adsabs.harvard.edu/abs/2015A&A...580A..65M}.
\end{barticle}
\endbibitem

\bibitem[\protect\citeauthoryear{{Morosan} \textit{et~al.}}{2017}]{mo17}
\begin{barticle}
\bauthor{\bsnm{{Morosan}}, \binits{D.E.}},
\bauthor{\bsnm{{Gallagher}}, \binits{P.T.}},
\bauthor{\bsnm{{Fallows}}, \binits{R.A.}},
\bauthor{\bsnm{{Reid}}, \binits{H.}},
\bauthor{\bsnm{{Mann}}, \binits{G.}},
\bauthor{\bsnm{{Bisi}}, \binits{M.M.}},
\bauthor{\bsnm{{Magdaleni{\'c}}}, \binits{J.}},
\bauthor{},
\bauthor{\bparticle{et.} \bsnm{al.}}:
\byear{2017},
\batitle{{The association of a J-burst with a solar jet}}.
\bjtitle{\aap}
\bvolume{606},
\bfpage{A81}.
\doiurl{10.1051/0004-6361/201629996}.
\adsurl{https://ui.adsabs.harvard.edu/abs/2017A&A...606A..81M}.
\end{barticle}
\endbibitem

\bibitem[\protect\citeauthoryear{{Morosan} \textit{et~al.}}{2019}]{mo19}
\begin{barticle}
\bauthor{\bsnm{{Morosan}}, \binits{D.E.}},
\bauthor{\bsnm{{Carley}}, \binits{E.P.}},
\bauthor{\bsnm{{Hayes}}, \binits{L.A.}},
\bauthor{\bsnm{{Murray}}, \binits{S.A.}},
\bauthor{\bsnm{{Zucca}}, \binits{P.}},
\bauthor{\bsnm{{Fallows}}, \binits{R.A.}},
\bauthor{\bsnm{{McCauley}}, \binits{J.}},
\bauthor{\bsnm{{Kilpua}}, \binits{E.K.J.}},
\bauthor{\bsnm{{Mann}}, \binits{G.}},
\bauthor{\bsnm{{Vocks}}, \binits{C.}},
\bauthor{\bsnm{{Gallagher}}, \binits{P.T.}}:
\byear{2019},
\batitle{{Multiple regions of shock-accelerated particles during a solar
  coronal mass ejection}}.
\bjtitle{Nature Astronomy}
\bvolume{3},
\bfpage{452}.
\doiurl{10.1038/s41550-019-0689-z}.
\adsurl{https://ui.adsabs.harvard.edu/abs/2019NatAs...3..452M}.
\end{barticle}
\endbibitem

\bibitem[\protect\citeauthoryear{{Mugundhan}
  \textit{et~al.}}{2018a}]{Mugundhan2018a}
\begin{barticle}
\bauthor{\bsnm{{Mugundhan}}, \binits{V.}},
\bauthor{\bsnm{{Ramesh}}, \binits{R.}},
\bauthor{\bsnm{{Kathiravan}}, \binits{C.}},
\bauthor{\bsnm{{Gireesh}}, \binits{G.V.S.}},
\bauthor{\bsnm{{Hegde}}, \binits{A.}}:
\byear{2018}a,
\batitle{{Spectropolarimetric Observations of Solar Noise Storms at Low
  Frequencies}}.
\bjtitle{\solphys}
\bvolume{293}(\bissue{3}),
\bfpage{41}.
\doiurl{10.1007/s11207-018-1260-2}.
\adsurl{https://ui.adsabs.harvard.edu/abs/2018SoPh..293...41M}.
\end{barticle}
\endbibitem

\bibitem[\protect\citeauthoryear{{Mugundhan}
  \textit{et~al.}}{2018b}]{Mugundhan2018b}
\begin{barticle}
\bauthor{\bsnm{{Mugundhan}}, \binits{V.}},
\bauthor{\bsnm{{Ramesh}}, \binits{R.}},
\bauthor{\bsnm{{Kathiravan}}, \binits{C.}},
\bauthor{\bsnm{{Gireesh}}, \binits{G.V.S.}},
\bauthor{\bsnm{{Kumari}}, \binits{A.}},
\bauthor{\bsnm{{Hariharan}}, \binits{K.}},
\bauthor{\bsnm{{Barve}}, \binits{I.V.}}:
\byear{2018}b,
\batitle{{The First Low-frequency Radio Observations of the Solar Corona on
  {\ensuremath{\approx}}200 km Long Interferometer Baseline}}.
\bjtitle{\apjl}
\bvolume{855}(\bissue{1}),
\bfpage{L8}.
\doiurl{10.3847/2041-8213/aaaf64}.
\adsurl{https://ui.adsabs.harvard.edu/abs/2018ApJ...855L...8M}.
\end{barticle}
\endbibitem

\bibitem[\protect\citeauthoryear{{Payne-Scott}}{1949}]{Payne1949}
\begin{barticle}
\bauthor{\bsnm{{Payne-Scott}}, \binits{R.}}:
\byear{1949},
\batitle{{Bursts of Solar Radiation at Metre Wavelengths}}.
\bjtitle{Australian Journal of Scientific Research A Physical Sciences}
\bvolume{2},
\bfpage{214}.
\doiurl{10.1071/PH490214}.
\adsurl{https://ui.adsabs.harvard.edu/abs/1949AuSRA...2..214P}.
\end{barticle}
\endbibitem

\bibitem[\protect\citeauthoryear{{Rahman}, {Cairns}, and
  {McCauley}}{2020}]{rahman20}
\begin{barticle}
\bauthor{\bsnm{{Rahman}}, \binits{M.M.}},
\bauthor{\bsnm{{Cairns}}, \binits{I.H.}},
\bauthor{\bsnm{{McCauley}}, \binits{P.I.}}:
\byear{2020},
\batitle{{Spectropolarimetric Imaging of Metric Type III Solar Radio Bursts}}.
\bjtitle{\solphys}
\bvolume{295}(\bissue{3}),
\bfpage{51}.
\doiurl{10.1007/s11207-020-01616-0}.
\adsurl{https://ui.adsabs.harvard.edu/abs/2020SoPh..295...51R}.
\end{barticle}
\endbibitem

\bibitem[\protect\citeauthoryear{{Ramesh}, {Mugundhan}, and
  {Prabhu}}{2020}]{ra20}
\begin{barticle}
\bauthor{\bsnm{{Ramesh}}, \binits{R.}},
\bauthor{\bsnm{{Mugundhan}}, \binits{V.}},
\bauthor{\bsnm{{Prabhu}}, \binits{K.}}:
\byear{2020},
\batitle{{New Evidence for Spatio-temporal Fragmentation in the Solar Flare
  Energy Release}}.
\bjtitle{\apjl}
\bvolume{889}(\bissue{1}),
\bfpage{L25}.
\doiurl{10.3847/2041-8213/ab6a9c}.
\adsurl{https://ui.adsabs.harvard.edu/abs/2020ApJ...889L..25R}.
\end{barticle}
\endbibitem

\bibitem[\protect\citeauthoryear{{Ramesh} \textit{et~al.}}{2008}]{Ramesh2008}
\begin{barticle}
\bauthor{\bsnm{{Ramesh}}, \binits{R.}},
\bauthor{\bsnm{{Kathiravan}}, \binits{C.}},
\bauthor{\bsnm{{Sundararajan}}, \binits{M.S.}},
\bauthor{\bsnm{{Barve}}, \binits{I.V.}},
\bauthor{\bsnm{{Sastry}}, \binits{C.V.}}:
\byear{2008},
\batitle{{A Low-Frequency (30 - 110 MHz) Antenna System for Observations of
  Polarized Radio Emission from the Solar Corona}}.
\bjtitle{\solphys}
\bvolume{253}(\bissue{1-2}),
\bfpage{319}.
\doiurl{10.1007/s11207-008-9272-y}.
\adsurl{https://ui.adsabs.harvard.edu/abs/2008SoPh..253..319R}.
\end{barticle}
\endbibitem

\bibitem[\protect\citeauthoryear{{Ramesh} \textit{et~al.}}{2013}]{Ramesh2013}
\begin{barticle}
\bauthor{\bsnm{{Ramesh}}, \binits{R.}},
\bauthor{\bsnm{{Sasikumar Raja}}, \binits{K.}},
\bauthor{\bsnm{{Kathiravan}}, \binits{C.}},
\bauthor{\bsnm{{Narayanan}}, \binits{A.S.}}:
\byear{2013},
\batitle{{Low-frequency Radio Observations of Picoflare Category Energy
  Releases in the Solar Atmosphere}}.
\bjtitle{\apj}
\bvolume{762}(\bissue{2}),
\bfpage{89}.
\doiurl{10.1088/0004-637X/762/2/89}.
\adsurl{https://ui.adsabs.harvard.edu/abs/2013ApJ...762...89R}.
\end{barticle}
\endbibitem

\bibitem[\protect\citeauthoryear{{Reid} and {Kontar}}{2017}]{re17}
\begin{barticle}
\bauthor{\bsnm{{Reid}}, \binits{H.A.S.}},
\bauthor{\bsnm{{Kontar}}, \binits{E.P.}}:
\byear{2017},
\batitle{{Imaging spectroscopy of type U and J solar radio bursts with LOFAR}}.
\bjtitle{\aap}
\bvolume{606},
\bfpage{A141}.
\doiurl{10.1051/0004-6361/201730701}.
\adsurl{https://ui.adsabs.harvard.edu/abs/2017A&A...606A.141R}.
\end{barticle}
\endbibitem

\bibitem[\protect\citeauthoryear{{Reid} and {Ratcliffe}}{2014}]{Reid2014}
\begin{barticle}
\bauthor{\bsnm{{Reid}}, \binits{H.A.S.}},
\bauthor{\bsnm{{Ratcliffe}}, \binits{H.}}:
\byear{2014},
\batitle{{A review of solar type III radio bursts}}.
\bjtitle{Research in Astronomy and Astrophysics}
\bvolume{14}(\bissue{7}),
\bfpage{773}.
\doiurl{10.1088/1674-4527/14/7/003}.
\adsurl{https://ui.adsabs.harvard.edu/abs/2014RAA....14..773R}.
\end{barticle}
\endbibitem

\bibitem[\protect\citeauthoryear{{Reiner}, {Fainberg}, and
  {Stone}}{1995}]{re95}
\begin{barticle}
\bauthor{\bsnm{{Reiner}}, \binits{M.J.}},
\bauthor{\bsnm{{Fainberg}}, \binits{J.}},
\bauthor{\bsnm{{Stone}}, \binits{R.G.}}:
\byear{1995},
\batitle{{Large-Scale Interplanetary Magnetic Field Configuration Revealed by
  Solar Radio Bursts}}.
\bjtitle{Science}
\bvolume{270},
\bfpage{461}.
\doiurl{10.1126/science.270.5235.461}.
\adsurl{1995Sci...270..461R}.
\end{barticle}
\endbibitem

\bibitem[\protect\citeauthoryear{{Robinson}}{1983}]{ro82}
\begin{barticle}
\bauthor{\bsnm{{Robinson}}, \binits{R.D.}}:
\byear{1983},
\batitle{{Scattering of radio waves in the solar corona}}.
\bjtitle{Proceedings of the Astronomical Society of Australia}
\bvolume{5}(\bissue{2}),
\bfpage{208}.
\doiurl{10.1017/S132335800001688X}.
\adsurl{https://ui.adsabs.harvard.edu/abs/1983PASAu...5..208R}.
\end{barticle}
\endbibitem

\bibitem[\protect\citeauthoryear{{Robishaw} and {Heiles}}{2021}]{Robishaw2021}
\begin{bbook}
\bauthor{\bsnm{{Robishaw}}, \binits{T.}},
\bauthor{\bsnm{{Heiles}}, \binits{C.}}:
\byear{2021},
In: \beditor{\bsnm{{Wolszczan}}, \binits{A.}} (ed.)
\bbtitle{{The Measurement of Polarization in Radio Astronomy}},
\bfpage{127}.
\doiurl{10.1142/9789811203770\_0006}.
\adsurl{https://ui.adsabs.harvard.edu/abs/2021hai1.book..127R}.
\end{bbook}
\endbibitem

\bibitem[\protect\citeauthoryear{{Saint-Hilaire}, {Vilmer}, and
  {Kerdraon}}{2013}]{sa13}
\begin{barticle}
\bauthor{\bsnm{{Saint-Hilaire}}, \binits{P.}},
\bauthor{\bsnm{{Vilmer}}, \binits{N.}},
\bauthor{\bsnm{{Kerdraon}}, \binits{A.}}:
\byear{2013},
\batitle{{A Decade of Solar Type III Radio Bursts Observed by the Nan{\c c}ay
  Radioheliograph 1998-2008}}.
\bjtitle{\apj}
\bvolume{762},
\bfpage{60}.
\doiurl{10.1088/0004-637X/762/1/60}.
\adsurl{2013ApJ...762...60S}.
\end{barticle}
\endbibitem

\bibitem[\protect\citeauthoryear{{Santandrea}
  \textit{et~al.}}{2013}]{Santandrea2013}
\begin{barticle}
\bauthor{\bsnm{{Santandrea}}, \binits{S.}},
\bauthor{\bsnm{{Gantois}}, \binits{K.}},
\bauthor{\bsnm{{Strauch}}, \binits{K.}},
\bauthor{\bsnm{{Teston}}, \binits{F.}},
\bauthor{\bsnm{{Tilmans}}, \binits{E.}},
\bauthor{\bsnm{{Baijot}}, \binits{C.}},
\bauthor{\bsnm{{Gerrits}}, \binits{D.}},
\bauthor{\bsnm{{De Groof}}, \binits{A.}},
\bauthor{\bsnm{{Schwehm}}, \binits{G.}},
\bauthor{\bsnm{{Zender}}, \binits{J.}}:
\byear{2013},
\batitle{{PROBA2: Mission and Spacecraft Overview}}.
\bjtitle{\solphys}
\bvolume{286}(\bissue{1}),
\bfpage{5}.
\doiurl{10.1007/s11207-013-0289-5}.
\adsurl{https://ui.adsabs.harvard.edu/abs/2013SoPh..286....5S}.
\end{barticle}
\endbibitem

\bibitem[\protect\citeauthoryear{{Sasikumar Raja}
  \textit{et~al.}}{2013}]{Sasi2013}
\begin{barticle}
\bauthor{\bsnm{{Sasikumar Raja}}, \binits{K.}},
\bauthor{\bsnm{{Kathiravan}}, \binits{C.}},
\bauthor{\bsnm{{Ramesh}}, \binits{R.}},
\bauthor{\bsnm{{Rajalingam}}, \binits{M.}},
\bauthor{\bsnm{{Barve}}, \binits{I.V.}}:
\byear{2013},
\batitle{{Design and Performance of a Low-frequency Cross-polarized
  Log-periodic Dipole Antenna}}.
\bjtitle{\apjs}
\bvolume{207}(\bissue{1}),
\bfpage{2}.
\doiurl{10.1088/0067-0049/207/1/2}.
\adsurl{https://ui.adsabs.harvard.edu/abs/2013ApJS..207....2S}.
\end{barticle}
\endbibitem

\bibitem[\protect\citeauthoryear{{Sastry}}{2009}]{Sastry2009}
\begin{barticle}
\bauthor{\bsnm{{Sastry}}, \binits{C.V.}}:
\byear{2009},
\batitle{{Polarization of the Thermal Radio Emission from Outer Solar Corona}}.
\bjtitle{\apj}
\bvolume{697}(\bissue{2}),
\bfpage{1934}.
\doiurl{10.1088/0004-637X/697/2/1934}.
\adsurl{https://ui.adsabs.harvard.edu/abs/2009ApJ...697.1934S}.
\end{barticle}
\endbibitem

\bibitem[\protect\citeauthoryear{{Seaton} \textit{et~al.}}{2013}]{se13}
\begin{barticle}
\bauthor{\bsnm{{Seaton}}, \binits{D.B.}},
\bauthor{\bsnm{{Berghmans}}, \binits{D.}},
\bauthor{\bsnm{{Nicula}}, \binits{B.}},
\bauthor{\bsnm{{Halain}}, \binits{J.-P.}},
\bauthor{\bsnm{{De Groof}}, \binits{A.}},
\bauthor{\bsnm{{Thibert}}, \binits{T.}},
\bauthor{\bsnm{{Bloomfield}}, \binits{D.S.}},
\bauthor{\bsnm{{Raftery}}, \binits{C.L.}},
\bauthor{\bsnm{{Gallagher}}, \binits{P.T.}},
\bauthor{\bsnm{{Auch{\`e}re}}, \binits{F.}},
\bauthor{\bsnm{{Defise}}, \binits{J.-M.}},
\bauthor{\bsnm{{D'Huys}}, \binits{E.}},
\bauthor{\bsnm{{Lecat}}, \binits{J.-H.}},
\bauthor{\bsnm{{Mazy}}, \binits{E.}},
\bauthor{\bsnm{{Rochus}}, \binits{P.}},
\bauthor{\bsnm{{Rossi}}, \binits{L.}},
\bauthor{\bsnm{{Sch{\"u}hle}}, \binits{U.}},
\bauthor{\bsnm{{Slemzin}}, \binits{V.}},
\bauthor{\bsnm{{Yalim}}, \binits{M.S.}},
\bauthor{\bsnm{{Zender}}, \binits{J.}}:
\byear{2013},
\batitle{{The SWAP EUV Imaging Telescope Part I: Instrument Overview and Pre-
  Flight Testing}}.
\bjtitle{\solphys}
\bvolume{286},
\bfpage{43}.
\doiurl{10.1007/s11207-012-0114-6}.
\adsurl{https://ui.adsabs.harvard.edu/abs/2013SoPh..286...43S}.
\end{barticle}
\endbibitem

\bibitem[\protect\citeauthoryear{{Spicer}, {Benz}, and
  {Huba}}{1982}]{Spicer1982}
\begin{barticle}
\bauthor{\bsnm{{Spicer}}, \binits{D.S.}},
\bauthor{\bsnm{{Benz}}, \binits{A.O.}},
\bauthor{\bsnm{{Huba}}, \binits{J.D.}}:
\byear{1982},
\batitle{{Solar type I noise storms and newly emerging magnetic flux}}.
\bjtitle{\aap}
\bvolume{105}(\bissue{2}),
\bfpage{221}.
\adsurl{https://ui.adsabs.harvard.edu/abs/1982A&A...105..221S}.
\end{barticle}
\endbibitem

\bibitem[\protect\citeauthoryear{{Stappers} \textit{et~al.}}{2011}]{st11}
\begin{barticle}
\bauthor{\bsnm{{Stappers}}, \binits{B.W.}},
\bauthor{\bsnm{{Hessels}}, \binits{J.W.T.}},
\bauthor{\bsnm{{Alexov}}, \binits{A.}},
\bauthor{\bsnm{{Anderson}}, \binits{K.}},
\bauthor{\bsnm{{Coenen}}, \binits{T.}},
\bauthor{\bsnm{{Hassall}}, \binits{T.}},
\bauthor{\bparticle{et.} \bsnm{al.}}:
\byear{2011},
\batitle{{Observing pulsars and fast transients with LOFAR}}.
\bjtitle{\aap}
\bvolume{530},
\bfpage{A80}.
\doiurl{10.1051/0004-6361/201116681}.
\adsurl{https://ui.adsabs.harvard.edu/abs/2011A&A...530A..80S}.
\end{barticle}
\endbibitem

\bibitem[\protect\citeauthoryear{{Stewart} and {Labrum}}{1972}]{Stewart1972}
\begin{barticle}
\bauthor{\bsnm{{Stewart}}, \binits{R.T.}},
\bauthor{\bsnm{{Labrum}}, \binits{N.R.}}:
\byear{1972},
\batitle{{Meter-wavelength observations of the solar radio burst storm of
  August 17 22, 1968}}.
\bjtitle{\solphys}
\bvolume{27}(\bissue{1}),
\bfpage{192}.
\doiurl{10.1007/BF00151783}.
\adsurl{https://ui.adsabs.harvard.edu/abs/1972SoPh...27..192S}.
\end{barticle}
\endbibitem

\bibitem[\protect\citeauthoryear{{Stone} and {Fainberg}}{1971}]{st71}
\begin{barticle}
\bauthor{\bsnm{{Stone}}, \binits{R.G.}},
\bauthor{\bsnm{{Fainberg}}, \binits{J.}}:
\byear{1971},
\batitle{{A U-Type Solar Radio Burst Originating in the Outer Corona}}.
\bjtitle{\solphys}
\bvolume{20},
\bfpage{106}.
\doiurl{10.1007/BF00146101}.
\adsurl{1971SoPh...20..106S}.
\end{barticle}
\endbibitem

\bibitem[\protect\citeauthoryear{{Suzuki} and {Dulk}}{1985}]{Suzuki1985}
\begin{bbook}
\bauthor{\bsnm{{Suzuki}}, \binits{S.}},
\bauthor{\bsnm{{Dulk}}, \binits{G.A.}}:
\byear{1985},
In: \beditor{\bsnm{{McLean}}, \binits{D.J.}},
\beditor{\bsnm{{Labrum}}, \binits{N.R.}} (eds.)
\bbtitle{{Bursts of type III and type V.}},
\bfpage{289}.
\adsurl{https://ui.adsabs.harvard.edu/abs/1985srph.book..289S}.
\end{bbook}
\endbibitem

\bibitem[\protect\citeauthoryear{{Swarup}, {Stone}, and
  {Maxwell}}{1960}]{Swarup1960}
\begin{barticle}
\bauthor{\bsnm{{Swarup}}, \binits{G.}},
\bauthor{\bsnm{{Stone}}, \binits{P.H.}},
\bauthor{\bsnm{{Maxwell}}, \binits{A.}}:
\byear{1960},
\batitle{{The Association of Solar Radio Bursts with Flares and Prominences.}}
\bjtitle{\apj}
\bvolume{131},
\bfpage{725}.
\doiurl{10.1086/146885}.
\adsurl{https://ui.adsabs.harvard.edu/abs/1960ApJ...131..725S}.
\end{barticle}
\endbibitem

\bibitem[\protect\citeauthoryear{Takakura}{1963}]{takakura1963}
\begin{barticle}
\bauthor{\bsnm{Takakura}, \binits{T.}}:
\byear{1963},
\batitle{Origin of solar radio type i bursts}.
\bjtitle{Publications of the Astronomical Society of Japan}
\bvolume{15},
\bfpage{462}.
\end{barticle}
\endbibitem

\bibitem[\protect\citeauthoryear{{Thejappa} and {Kundu}}{1991}]{Thejappa1991}
\begin{barticle}
\bauthor{\bsnm{{Thejappa}}, \binits{G.}},
\bauthor{\bsnm{{Kundu}}, \binits{M.R.}}:
\byear{1991},
\batitle{{New Observations of Solar Noise Storm Radiation at Decameter
  Wavelengths}}.
\bjtitle{\solphys}
\bvolume{132}(\bissue{1}),
\bfpage{155}.
\doiurl{10.1007/BF00159136}.
\adsurl{https://ui.adsabs.harvard.edu/abs/1991SoPh..132..155T}.
\end{barticle}
\endbibitem

\bibitem[\protect\citeauthoryear{{Tingay} \textit{et~al.}}{2013}]{mwa13}
\begin{barticle}
\bauthor{\bsnm{{Tingay}}, \binits{S.J.}},
\bauthor{\bsnm{{Goeke}}, \binits{R.}},
\bauthor{\bsnm{{Bowman}}, \binits{J.D.}},
\bauthor{\bsnm{{Emrich}}, \binits{D.}},
\bauthor{\bsnm{{Ord}}, \binits{S.M.}},
\bauthor{\bsnm{{Mitchell}}, \binits{D.A.}},
\bauthor{\bsnm{{Morales}}, \binits{M.F.}},
\bauthor{\bsnm{{Booler}}, \binits{T.}},
\bauthor{\bparticle{et.} \bsnm{al.}}:
\byear{2013},
\batitle{{The Murchison Widefield Array: The Square Kilometre Array Precursor
  at Low Radio Frequencies}}.
\bjtitle{\pasa}
\bvolume{30},
\bfpage{e007}.
\doiurl{10.1017/pasa.2012.007}.
\adsurl{https://ui.adsabs.harvard.edu/abs/2013PASA...30....7T}.
\end{barticle}
\endbibitem

\bibitem[\protect\citeauthoryear{{van Haarlem} \textit{et~al.}}{2013}]{lofar13}
\begin{barticle}
\bauthor{\bsnm{{van Haarlem}}, \binits{M.P.}},
\bauthor{\bsnm{{Wise}}, \binits{M.W.}},
\bauthor{\bsnm{{Gunst}}, \binits{A.W.}},
\bauthor{\bsnm{{Heald}}, \binits{G.}},
\bauthor{\bsnm{{McKean}}, \binits{J.P.}},
\bauthor{\bsnm{{Hessels}}, \binits{J.W.T.}},
\bauthor{\bparticle{et.} \bsnm{al.}}:
\byear{2013},
\batitle{{LOFAR: The LOw-Frequency ARray}}.
\bjtitle{\aap}
\bvolume{556},
\bfpage{A2}.
\doiurl{10.1051/0004-6361/201220873}.
\adsurl{2013A\%26A...556A...2V}.
\end{barticle}
\endbibitem

\bibitem[\protect\citeauthoryear{{Wentzel}}{1984}]{wentzel84}
\begin{barticle}
\bauthor{\bsnm{{Wentzel}}, \binits{D.G.}}:
\byear{1984},
\batitle{{Polarization of Fundamental Type-Iii Radio Bursts}}.
\bjtitle{\solphys}
\bvolume{90}(\bissue{1}),
\bfpage{139}.
\doiurl{10.1007/BF00153791}.
\adsurl{https://ui.adsabs.harvard.edu/abs/1984SoPh...90..139W}.
\end{barticle}
\endbibitem

\bibitem[\protect\citeauthoryear{{Wild}}{1950}]{wild50}
\begin{barticle}
\bauthor{\bsnm{{Wild}}, \binits{J.P.}}:
\byear{1950},
\batitle{{Observations of the Spectrum of High-Intensity Solar Radiation at
  Metre Wavelengths. III. Isolated Bursts}}.
\bjtitle{Australian Journal of Scientific Research A Physical Sciences}
\bvolume{3},
\bfpage{541}.
\adsurl{1950AuSRA...3..541W}.
\end{barticle}
\endbibitem

\bibitem[\protect\citeauthoryear{{Wild}}{1951}]{Wild1951}
\begin{barticle}
\bauthor{\bsnm{{Wild}}, \binits{J.P.}}:
\byear{1951},
\batitle{{Observations of the Spectrum of High-Intensity Solar Radiation at
  Metre Wavelengths. IV. Enhanced Radiation}}.
\bjtitle{Australian Journal of Scientific Research A Physical Sciences}
\bvolume{4},
\bfpage{36}.
\doiurl{10.1071/PH510036}.
\adsurl{https://ui.adsabs.harvard.edu/abs/1951AuSRA...4...36W}.
\end{barticle}
\endbibitem

\bibitem[\protect\citeauthoryear{{Wild}}{1963}]{wild63}
\begin{botherref}
\oauthor{\bsnm{{Wild}}, \binits{J.P.}}:
1963,
{The radio emission of the sun}.
\textit{Radiotekhnika},
1.
\adsurl{1963rat..conf....1W}.
\end{botherref}
\endbibitem

\bibitem[\protect\citeauthoryear{{Wild}}{1967}]{wi67}
\begin{barticle}
\bauthor{\bsnm{{Wild}}, \binits{J.P.}}:
\byear{1967},
\batitle{{The radioheliograph and the radio astronomy programme of the Culgoora
  Observatory}}.
\bjtitle{Proceedings of the Astronomical Society of Australia}
\bvolume{1},
\bfpage{38}.
\adsurl{1967PASAu...1...38W}.
\end{barticle}
\endbibitem

\bibitem[\protect\citeauthoryear{{Yu} \textit{et~al.}}{2019}]{Yu2019}
\begin{bchapter}
\bauthor{\bsnm{{Yu}}, \binits{S.}},
\bauthor{\bsnm{{Chen}}, \binits{B.}},
\bauthor{\bsnm{{Bastian}}, \binits{T.}},
\bauthor{\bsnm{{Gary}}, \binits{D.E.}}:
\byear{2019},
\bctitle{{Imaging Spectroscopic Observations of Type I Noise Storms with
  Ultrahigh Temporal and Spectral Resolution}}.
In: \bbtitle{AGU Fall Meeting Abstracts}
\bseriesno{2019},
\bfpage{SH23C}.
\adsurl{https://ui.adsabs.harvard.edu/abs/2019AGUFMSH23C3336Y}.
\end{bchapter}
\endbibitem

\bibitem[\protect\citeauthoryear{}{}]{mu18}
\begin{botherref}
.
\end{botherref}
\endbibitem

\end{thebibliography}

\end{article} 

\end{document}